\documentclass[a4paper,11pt]{article}
\pdfoutput=1
\usepackage{graphicx,rotating,hyperref,slashed,amsmath,xcolor,amssymb,amsfonts,colortbl,cite, subfigure,float}
\makeatletter 
\usepackage{graphics}
\usepackage{graphicx} 
\graphicspath{{plots/}}

\hypersetup{colorlinks,bookmarksopen,bookmarksnumbered,
linkcolor=blus,pdfstartview=FitH,urlcolor=rossos,citecolor=verde}
\allowdisplaybreaks

\def\lsim{\mathrel{\rlap{\lower3pt\hbox{\hskip0pt$\sim$}}
   \raise1pt\hbox{$<$}}}         
\def\gsim{\mathrel{\rlap{\lower4pt\hbox{\hskip1pt$\sim$}}
   \raise1pt\hbox{$>$}}}         

 \newcommand{\sfootnote}[1]{} 
\definecolor{bluc}{cmyk}{1,1,0,0.1}
\definecolor{rossoCP3}{cmyk}{0,.88,.77,.40}
\definecolor{rosso}{cmyk}{0,1,1,0.4}
\definecolor{rossos}{cmyk}{0,1,1,0.55}
\definecolor{rossoc}{cmyk}{0,1,1,0.2}
\definecolor{verdes}{cmyk}{0.92,0,0.59,0.4}

\hypersetup{colorlinks, bookmarksopen, bookmarksnumbered,
citecolor=verdes, linkcolor=bluc, pdfstartview=FitH, urlcolor=rossos}

\newcommand{\mio}[1]{}

\definecolor{Gray}{gray}{0.95}

\newcommand{\e}{{\bf e}}

\usepackage{multicol}
\usepackage{color}
\definecolor{rosso}{cmyk}{0,1,1,0.4}
\definecolor{rossos}{cmyk}{0,1,1,0.55}
\definecolor{rossoc}{cmyk}{0,1,1,0.2}
\definecolor{blu}{cmyk}{1,1,0,0.3}
\definecolor{blus}{cmyk}{1,1,0,0.6}
\definecolor{bluc}{cmyk}{1,1,0,0.1}
\definecolor{verde}{cmyk}{0.92,0,0.59,0.25}
\definecolor{verdec}{cmyk}{0.92,0,0.59,0.15}
\definecolor{verdes}{cmyk}{0.92,0,0.59,0.4}

\renewcommand\&{&}

\def\circa#1{\,\raise.3ex\hbox{$#1$\kern-.75em\lower1ex\hbox{$\sim$}}\,}

\newcommand{\beq}{\begin{equation}}
\newcommand{\eeq}{\end{equation}}

\newcommand{\bea}{\begin{eqnarray}}
\newcommand{\eea}{\end{eqnarray}}
\newcommand{\be}{\begin{equation}}
\newcommand{\ee}{\end{equation}}
\newfam\rsfsfam
\def\mathscr#1{{\fam\rsfsfam\relax#1}}

\def\circa#1{\,\raise.3ex\hbox{$#1$\kern-.75em\lower1ex\hbox{$\sim$}}\,}
\makeatletter

\def\hhref#1{\href{http://arxiv.org/abs/#1}{arXiv:#1}} 

\newcommand{\doi}[1]{\href{http://dx.doi.org/#1}{[doi]}}

\def\hhref#1{\href{http://arxiv.org/abs/#1}{arXiv:#1}} 
 
\def\art{\@ifnextchar[{\eart}{\oart}}
\def\eart[#1]#2#3#4#5#6{{\rm #2}, {\em #3 \bf #4} {\rm (#6) #5} ({\em #1})}

\def\article{\@ifnextchar[{\earticle}{\oarticle}}
\def\oarticle#1#2#3#4#5#6{{\rm #1}, {\em ``#6''}, {\rm #2 #3 (#5) #4}}
\def\earticle[#1]#2#3#4#5#6#7{{\rm #2}, {\em ``#7''}, {\rm #3 #4 (#6) #5}  [\hhref{#1}]}
\def\hepart[#1]#2{{\rm #2, \em#1}}
\def\heparticle[#1]#2#3{#2, {\em ``#3''} [\hhref{#1}]}

\newcounter{alphaequation}[equation]
\def\thealphaequation{\theequation\hbox to
0.6em{\hfil\alph{alphaequation}\hfil}}
\def\eqnsystem#1{
\def\@eqnnum{{\rm (\thealphaequation)}}
\def\@@eqncr{\let\@tempa\relax \ifcase\@eqcnt \def\@tempa{& & &} \or
  \def\@tempa{& &}\or \def\@tempa{&}\fi\@tempa
  \if@eqnsw\@eqnnum\refstepcounter{alphaequation}\fi
\global\@eqnswtrue\global\@eqcnt=0\cr}
\refstepcounter{equation} \let\@currentlabel\theequation \def\@tempb{#1}
\ifx\@tempb\empty\else\label{#1}\fi
\refstepcounter{alphaequation}
\let\@currentlabel\thealphaequation
\global\@eqnswtrue\global\@eqcnt=0 \tabskip\@centering\let\\=\@eqncr
$$\halign to \displaywidth\bgroup \@eqnsel\hskip\@centering
$\displaystyle\tabskip\z@{##}$&\global\@eqcnt\@ne
\hskip2\arraycolsep\hfil${##}$\hfil& \global\@eqcnt\tw@\hskip2\arraycolsep
$\displaystyle\tabskip\z@{##}$\hfil
\tabskip\@centering&\llap{##}\tabskip\z@\cr}
\def\endeqnsystem{\@@eqncr\egroup$$\global\@ignoretrue} \makeatother

\definecolor{fiorentina}{rgb}{.5,0,.5}

\begin{document}

\vspace{1truecm}
 \begin{center}
\boldmath

{\textbf{\Large 
Probing a stationary  non-Gaussian background \\ \vskip0.1cm of stochastic gravitational waves  with   pulsar timing arrays}}
\unboldmath
\end{center}
\unboldmath

\vspace{-0.2cm}

\begin{center}
\vspace{0.1truecm}
\date\today
{\bf Cari Powell, \, Gianmassimo Tasinato}
 \\[4mm]
{\it   Department of Physics, Swansea University, Swansea, SA2 8PP, United Kingdom
}\\[1mm]
\end{center}

\vspace{0.2cm}

\begin{abstract}
\noindent
{We introduce  the concept of stationary graviton non-Gaussianity (nG), an observable that can
be probed in terms of 3-point correlation functions of a stochastic
gravitational wave (GW) background. 
When evaluated in momentum space, stationary nG corresponds to folded bispectra of graviton nG.
 We determine 3-point  overlap functions 
for testing stationary  nG  with  pulsar timing array GW experiments, and we  obtain the
 corresponding optimal signal-to-noise ratio.
  For the first time, we consider 3-point overlap functions  including scalar graviton polarizations (which can be motivated
  in theories of modified gravity); moreover, we also calculate  3-point overlap functions for correlating pulsar timing array  with ground based
  GW detectors. 
  The value of the optimal signal-to-noise ratio  depends on the number and position of monitored pulsars. We build  geometrical quantities  characterizing how such ratio depends on the pulsar system under consideration, and   we  evaluate these  geometrical parameters     using  data from      the IPTA collaboration. We quantitatively show how monitoring a large number of pulsars can increase the  signal-to-noise ratio associated       with measurements of stationary graviton nG.}
\end{abstract}

\section{Introduction}\label{intro}
  After the direct detection of GWs from merging black hole and neutron star binaries, one of the
next challenges  for GW experiments is the measurement  of  a stochastic gravitational wave background 
(SGWB).
 A theoretical characterization of the properties of the SGWB  is 
essential for designing  observables aimed to distinguish among  different sources. 
Reviews
of astrophysical and cosmological sources for a SGWB measurable with GW experiments can be found e.g. in \cite{Allen:1996vm,Maggiore:2018sht,Bartolo:2016ami,Caprini:2018mtu,Regimbau:2011rp}.
 If a SGWB will be eventually  detected,  a natural question is whether it is possible to  disentangle its different contributions from astrophysical and/or  cosmological  sources.

 If a SGWB has cosmological origin, its spectrum can be characterized by  specific properties: the frequency dependence of its energy
 density profile can be more complex than the typical power-law that characterise astrophysical backgrounds. (See e.g. the recent \cite{Caprini:2019pxz} for an accurate 
tool  for distinguishing among different frequency profiles with LISA experiment.) Depending on the production mechanisms, it can be characterized by a large,  intrinsic 
graviton (also called tensor)  non-Gaussianity (nG) (see e.g. \cite{Bartolo:2018qqn} for an analysis and review
of tensor nG from cosmological inflation~\footnote{Also astrophysical backgrounds
can be non-Gaussian, when sources of GWs are at the verge of being individually detected: the
kind of nG is  different from the one discussed here, and requires
dedicated studies \cite{Drasco:2002yd,Seto:2009ju,Seto:2008xr,Racine:2007gv,Martellini:2014xia}.}).  Although GWs produced by early universe mechanisms can be non-Gaussian, any signal detected at  frequency scales of GW
 experiments is usually considered to be Gaussian,  for various related reasons \cite{Adshead:2009bz,Bartolo:2018evs,Bartolo:2018rku,Allen:1996vm}. One reason
 (as explained  in  \cite{Allen:1996vm,Adshead:2009bz}) 
  is that any higher order, connected correlation of signals detected with GW experiments typically
  involves angular integrations of  contributions from many different, causally disconnected patches of the sky.    By the central limit theorem, 
  such linear superposition of signals from  different directions tend to suppress any existing nG in GWs originating from each independent
  patch. Other  more concrete reasons, as
  spelled out in full detail in the recent works \cite{Bartolo:2018evs,Bartolo:2018rku} are as follows:
     on their  way through large cosmological distances from source
  to detection, GWs can collect random phases induced by long-wavelength matter fluctuations, which tend to suppress  existing non-Gaussian phase correlations
  among GW signals. Moreover, due to the finite time of  measurement, GW momenta can not be resolved perfectly, and such uncertainty   again suppresses non-Gaussian effects when measuring higher order correlators~\footnote{   Possible ways out to these negative conclusions have been proposed, involving  measurements of quantities   only indirectly sensitive to graviton nG:
 the  
  quadrupolar anisotropy of the SGWB power spectrum, an observable which depends on the squeezed limit of tensor non-Gaussian correlation functions \cite{Dimastrogiovanni:2019bfl,Ricciardone:2017kre,Ozsoy:2019slf}; and higher-order correlations among spatial anisotropies in the distribution function of the GW energy density \cite{Bartolo:2019oiq}.}.

\smallskip

A common feature of the cases studied so far, and partly at the root of the problems mentioned above, is that the corresponding GW signal 3-point function is not stationary: the value of the 3-point correlator of GW signal
evaluated 
at equal time (say $t$) depends on the value of $t$. In this work, to overcome this problem, we introduce and characterize the concept of {\it stationary graviton non-Gaussianity}.
   It is characterized by higher-order
    correlators with two important porperties: they  are invariant under time translation symmetry, and (as a consequence)  they   select GWs propagating along a common direction.
    Such features
     eliminate possible phase differences  accumulating along the way GW travel from source to detection. They
     can then allow one to  avoid the previous problems, making stationary graviton nG an observable that can be potentially 
   probed  by measuring 3-point functions of the SGWB with GW experiments.  When evaluated in momentum space, stationary nG
     corresponds to a folded (also dubbed flattened)   
    shape of tensor nG \footnote{The work \cite{Bartolo:2018evs} already pointed out that 3-point functions of GWs whose momenta are
    accurately aligned can avoid decorrelation effects.}.  In Section \ref{statio-nonG},  we first characterize general properties of this category 
    of non-Gaussian, stationary signals, and  explain why they have the opportunity to avoid the 
     problems investigated in  \cite{Bartolo:2018evs,Bartolo:2018rku}. We  then 
      discuss prospects
    to detect stationary non-Gaussianity with pulsar timing arrays (PTA).
    
        \smallskip
        
Besides interferometers, another promising tool for detect SGWBs is based on observations of time residuals from large arrays of pulsars, which can detect the passage of  GWs by tiny changes
in  their precisely measured periods.  They can detect GWs at  small frequencies of around $10^{-7} - 10^{-9}$ Hz. Several  collaborations are studying pulsar data set in order
to detect GW signals --  EPTA \cite{Kramer:2013kea}, NANOGrav \cite{McLaughlin:2013ira}, PPTA \cite{Manchester:2012za} -- and  data are collected in an international collaboration called IPTA \cite{Verbiest:2016vem} which is currently monitoring 49 pulsars. In the relatively  near future, SKA will considerably 
increase the number of monitored  pulsars and the accuracy of measurements, see e.g. \cite{Kramer:2004rwa}.  Theoretical studies
 of how the response of a PTA system to a  SGWB have been started
 decades ago by the work of  Hellings and Downs \cite{Hellings:1983fr}. A more recent,   detailed
analysis of  optimal signal-to-noise ratio and detectability prospects for a SGWB can be found in 
\cite{Anholm:2008wy}.  Reviews can be found in
\cite{Janssen:2014dka,Moore:2014lga,Maggiore:2018sht}. See also \cite{Tsuneto:2018tif} for a  study of overlap functions for 3-point non-Gaussian
correlators with PTA (we will discuss in footnote \ref{foot-comp}  the differences between  \cite{Tsuneto:2018tif} and our work). 
Discussions on tests of deviations from  General Relativity with PTAs can be found e.g. in \cite{Yunes:2013dva,Chamberlin:2011ev,Cornish:2017oic}.

 In Section \ref{sec_PTAoverlap}, we study overlap functions for PTA systems  associated with   stationary graviton  nG. We do so in various different cases. We first consider 3-point overlap functions for PTA data aimed to detect correlations among  
 spin-2 tensor modes of General Relativity. We then study 3-point overlap functions including scalar excitations (the `breathing mode' or transverse scalar graviton polarization), that are motivated by modified theories of gravity. We finally study correlations among different GW experiments (PTA and ground based interferometers),
  motivated by the fact 3-point functions associated with folded nG  can  correlate 
    signals with  very distinct   frequencies.
    
       Armed with these results, in Section \ref{sec-optimal} we determine the expression for the
  optimal signal-to-noise ratio to detect stationary nG  in the SGWB. We
   investigate 
  how the number of monitored pulsars and the geometry of the PTA system determine the optimal SGWB.  We compute some of the key geometrical
  quantities characterizing the optimal SNR with data from IPTA collaboration. Our results give a quantitative indication that  monitoring a large number of pulsars can increase the  signal-to-noise ratio associated       with measurements of stationary graviton nG.


\section{Characterization  of a  stationary non-Gaussian SGWB}\label{statio-nonG}

In this Section we discuss necessary conditions 
  to make  graviton non-Gaussianity   an observable that can be directly probed in terms of 
 $3$-point correlation functions of a SGWB.
  We assume that the SGWB  background is {\it stationary}, meaning that
all correlators are time translationally invariant. If they were not,   
destructive interference  effects are expected to set them to zero.
 Additionally, we  shall also assume that parity is conserved, and that the background 
geometry preserves 3-dimensional  spatial isotropy. 

\smallskip

The GW spin-2 tensor mode in transverse-traceless gauge  is expanded in Fourier modes  as 

\be \label{sig_fourier}
h_{ab} (t,\vec x)\,=\,\sum_\lambda \int_{-\infty}^\infty  d f \int d^2 \hat n\,e^{-2\pi\,i\,f\,\hat n\,\vec x }
\,
e^{2\pi\,i\,f\,t}\,
{\bf e}_{ab}^{(\lambda)}(\hat n)\,
 h_{\lambda}(f,\,\hat n)\,,
\ee
with $f$ the GW frequency, and $\hat n$ the unit vector corresponding to the GW direction.
 The product $2 \pi\,f\,\hat n$ corresponds to the 3-momentum of the GW. 
 The  condition $ h_{\lambda}(f,\,\hat n)= h^*_{\lambda}(-f,\,\hat n)$ ensures
   that the function  $h_{ab}(t,\vec x)$ is real. The sum runs
 over chirality index $\lambda=L, R$, and ${\bf e}_{ab}^{(\lambda)}(\hat n)$ denotes the polarization tensor: see Appendix \ref{app:conventions} for our conventions on these
 quantities.
In
  \eqref{sig_fourier}
we integrate over positive as well as negative frequencies $f$, so to maintain a concise expression \cite{Allen:1997ad}. 

\smallskip

Being the SGWB  by hypothesis stationary, 
  all correlators depend 
  on time differences only. In other words, correlators in real space as 
\be
\langle h_{a_1b_1} (t_1,\vec x_1)\dots  h_{a_n b_n} (t_n,\vec x_n)\rangle
\ee
depend only on $t_1-t_n$ for each $n$ and are invariant under time translation. This condition
is easily  achieved for the case of  2-point  correlation functions.  Assuming to correlate two modes with 
frequencies $f_{1,2}$ of opposite signs (say $f_1>0$),
  the
2-point correlator in Fourier space has the standard structure

\bea
 \label{builcor3a1}
\langle h_{\lambda_1}(f_1,\,\hat n_1) h_{\lambda_2}(f_2,\,\hat n_2) \rangle&=&
\,\delta^{(3)}\left(f_1\, \hat n_1+f_2 \,\hat n_2 \right)\,\delta^{\lambda_1 \lambda_2}
\,{ P}(f_1)\,,
\\
&=&\delta\left(f_1+f_2\right)
\,
\delta^{(2)}\left( \hat n_2-\hat n_3 \right)\,\delta^{\lambda_1 \lambda_2}
\,{ P}(f_1)\,,
 \label{builcor3a}
\eea
with $P(f)$ the power spectrum  depending on frequency.
In the second line, we used the fact that 
 the 3-dimensional $\delta$-function implies the condition $f_1 \hat n_1\,=\,-f_2 \hat n_2$. Taking the square
of this expression one gets $f_1^2\,=\,f_2^2\,\,\Rightarrow\,\,f_1\,=\,-f_2$ (recall that we are working
with  positive
as well as negative frequencies) and hence $ \hat n_1\,=\, \hat n_2$. 

The $\delta$-functions make the correlator in eq \eqref{builcor3a}   isotropic (the waves come from the same direction) and stationary. Stationarity is 
evident in
%
 the 2-point  correlator in real space: 
\bea
\langle h_{a_1b_1} (t_1,\vec x_1) h_{a_2b_2} (t_2,\vec x_2)  \rangle
&=&\sum_{\lambda_1 \lambda_2 }  \int_{-\infty}^\infty  d f_1  d f_2    \int d^2 \hat n_1 \, d^2 \hat n_2\, \nonumber
\\
&\times& e^{2\pi\,i\,f_1\,t_1}\, e^{2\pi\,i\,f_2\,t_2}\,\delta\left(f_1+f_2\right)\,\delta^{(2)}\left( \hat n_2-\hat n_3 \right)
\nonumber
\\
&\times&\e_{a_1b_1}^{(\lambda_1)}(\hat n_1)\,\e_{a_2b_2}^{(\lambda_2)}(\hat n_2)\,\,\delta^{\lambda_1 \lambda_2}\,
{ P}(f_1)
\\
&=&
\sum_{\lambda_1  }  \int_{-\infty}^\infty  d f_1      \int d^2 \hat n_1\, 
\,e^{2\pi\,i\,f_1\,\left(t_1-t_2 \right)}
\e_{a_1b_1}^{(\lambda_1)}(\hat n_1)\,\e_{a_2b_2}^{(\lambda_1)}(\hat n_1)\,\,
{ P}(f_1)\,.
\label{corrsta1a}
\eea
The previous expression   is time-translationally invariant,
 since it 
depends only on the time difference $(t_2-t_1)$ appearing 
in the exponential term of eq \eqref{corrsta1a}.
 
\bigskip

What about higher-order, connected $n$-point functions? The crucial feature of expression \eqref{builcor3a}
which leads to  time-translation invariance in real space  is the presence of the $\delta$-function in frequencies, $\delta(f_1+f_2)$.

 We
then  
{\it postulate}  that the same property holds for  the 3-point function in Fourier
space, and write the Ansatz 
%
%
%
%
%
%
 \be \label{builcor1}
\langle h_{\lambda_1}(f_1,\,\hat n_1) h_{\lambda_2}(f_2,\,\hat n_2) h_{\lambda_3}(f_3,\,\hat n_3) \rangle\,=\,\delta\left(f_1+f_2+f_3\right)
\,\langle h_{\lambda_1}(f_1,\,\hat n_1) h_{\lambda_2}(f_2,\,\hat n_2) h_{\lambda_3}(f_3,\,\hat n_3) \rangle_{\rm st.}
\ee
where the label ${\rm st}$ means stationary. 
A non-vanishing 3-pt correlator with this property characterizes what we shall call {\it stationary graviton 
non-Gaussianity} (nG). 
Indeed,
 substituting the decomposition   \eqref{builcor1} in the three point correlator
  in coordinate space, and using   eq \eqref{sig_fourier} to express correlators, we obtain the expression (we integrate over $f_3$ and use the $\delta$-function appearing in eq \eqref{builcor1})
\bea
\langle h_{a_1b_1} (t_1,\vec x_1) h_{a_2b_2} (t_2,\vec x_2)  h_{a_3 b_3} (t_3,\vec x_3)\rangle
&=&\sum_{\lambda_1 \lambda_2 \lambda_3}  \int_{-\infty}^\infty  d f_1  d f_2    \int d^2 \hat n_1 \, d^2 \hat n_2\, d^2 \hat n_3 
\nonumber
\\&&
\hskip-2cm
\times\, e^{2\pi\,i\,f_1\,\left(t_1-t_3\right)}\,
e^{2\pi\,i\,f_2\,\left(t_2-t_3\right)}\,e^{-2\pi\,i\,f_1\,\left(\hat n_1\,\vec x_1- \hat n_3\,\vec x_3\right)}\,e^{-2\pi\,i\,f_2\,\left(\hat n_2\,\vec x_2- \hat n_3\,\vec x_3\right)}
\nonumber
\\&&
\hskip-2cm
\times\, \e_{a_1b_1}^{(\lambda_1)}(\hat n_1)\,\e_{a_2b_2}^{(\lambda_2)}(\hat n_2)\,\,\e_{a_3b_3}^{(\lambda_3)}(\hat n_3)\,
\nonumber
\\&&
\hskip-2cm
\times\, 
\langle h_{\lambda_1}(f_1,\,\hat n_1) h_{\lambda_2}(f_2,\,\hat n_2) h^*_{\lambda_3}(f_1+f_2,\,\hat n_3) \rangle_{\rm st}
\,.
\label{corrsta1}
\eea
Since it depends only on time differences, this correlator is time translationally invariant, as desired, hence the 3-point function is stationary. 

\smallskip

At this level, our stationary Ansatz \eqref{builcor1} is purely phenomenological, being it introduced to realize the stationary condition \eqref{corrsta1}. But
it is not difficult  to 
 characterise  the 3-point function in Fourier space. Indeed,
the statistical  isotropy of the fluctuations also  requires that the   Fourier space correlator \eqref{builcor1}
   is proportional to the three dimensional   $\delta-$function acting on the three momenta, meaning that the vectors $f_i \hat n_i$ form
 a closed triangle (the label ${\rm iso}$ means isotropic): 
   \be \label{builcor2}
   \langle h_{\lambda_1}(f_1,\,\hat n_1) h_{\lambda_2}(f_2,\,\hat n_2) h_{\lambda_3}(f_3,\,\hat n_3) \rangle\,=\,
   \delta^{(3)}(f_1\,\hat n_1+f_2\,\hat n_2+f_3\,\hat n_3)\,
     \langle h_{\lambda_1}(f_1,\,\hat n_1) h_{\lambda_2}(f_2,\,\hat n_2) h_{\lambda_3}(f_3,\,\hat n_3) \rangle_{iso}
\,.
   \ee

\begin{figure}[H]
\begin{center}
\includegraphics[width = 0.7 \textwidth]{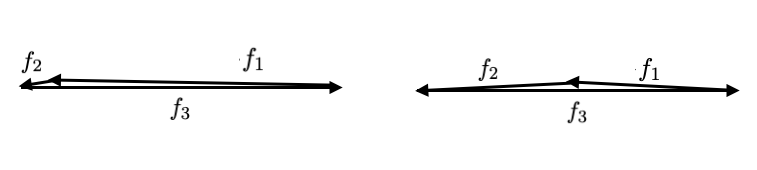}
 \caption{\it 
Representation of  folded (also called flattened) triangles. Here $f_{1,2}>0$, $f_3<0$.  The triangle sides are intended to be  superimposed, with vanishing
angles between the sides $(f_1,  f_3)$ and $(f_2, f_3)$. The side lengths can be 
very different (e.g. $f_1\sim |f_3|\gg f_2$,  left plot) or 
comparable in size (e.g. $f_1\sim f_2\sim |f_3|/2$, right plot).}
\label{fig:folded}
\end{center} 
\end{figure}

 We now show that, taken together with the condition \eqref{builcor2},  the stationarity condition of eq   \eqref{builcor1} selects only `folded' (also called flattened)  triangles in moment space, whose  sides are exactly superimposed.
 Since the sum over frequencies vanish for the stationarity condition, one of the two frequencies has
 sign opposite to the other one: without lack of generality, we can take $f_{1,2}>0$ and $f_3<0$. 
 We write the two conditions \eqref{builcor1} and \eqref{builcor2}  as
 \bea
 \label{cond1}
 f_3\hat n_3&=&-f_1 \hat n_1-f_2 \hat n_2\,,
 \\
  \label{cond2}
 f_3 &=&-f_1  -f_2\,.
 \eea
 Taking the square of both sides of \eqref{cond1}, we get the condition
  \bea
 \label{cond1a}
 f_3^2&=&f_1^2 +f_2^2 +2 \,f_1  f_2\, \hat n_1\cdot  \hat n_2
\,.
\eea 
Consider finite, non-vanishing values for $f_i$:
the only way to make eq \eqref{cond1a} compatible with  the square of both sides of eq \eqref{cond2} is to require $ \hat n_1 \cdot \hat n_2\,=\,1$. Contracting eq \eqref{cond1} with $\hat n_1$, 
and using this result,
we obtain
\bea
 \label{cond1b}
 f_3\,\hat n_3\cdot \hat n_1&=&-f_1 -f_2 \,,
 \eea
which is compatible with eq \eqref{cond2} only if $ \hat n_1 \cdot \hat n_3\,=\,1$. Hence the
condition of stationarity is equivalent  to consider folded triangles in momentum space, with  superimposed sides.
In other words, $\hat n_i \cdot \hat n_j\,=\,1$ for each $\hat n_i$, and the
 directions characterizing the GW modes entering
  the three-point  correlator lie on the same line. 
 See Fig \ref{fig:folded} for a graphical representation
of examples of folded triangles in momentum space, corresponding to folded nG.

To summarize, the condition of stationarity requires that the two and three point functions of tensor modes in Fourier space read, if   none of the $f_i$ is vanishing small~\footnote{In the squeezed case (one of the $f_i$ vanishes) then the condition of
stationarity
 it is not  necessarily associated  with the condition of superimposed   triangle sides. The consequences of squeezed configurations for modulating the tensor power spectrum have
been recently investigated in \cite{Dimastrogiovanni:2019bfl}.}:

\bea
 \label{builcor3a}
\langle h_{\lambda_1}(f_1,\,\hat n_1) h_{\lambda_2}(f_2,\,\hat n_2) \rangle&=&\delta\left(f_1+f_2\right)
\,\delta^{(2)}\left( \hat n_1-\hat n_2 \right)\,\delta^{\lambda_1 \lambda_2}
\,{ P}(f_1)\,,
\\
\langle h_{\lambda_1}(f_1,\,\hat n_1) h_{\lambda_2}(f_2,\,\hat n_2) h_{\lambda_3}(f_3,\,\hat n_3) \rangle&=&\delta\left(f_1+f_2+f_3\right)
\,\delta^{(2)}\left( \hat n_1-\hat n_3 \right)
\,\delta^{(2)}\left( \hat n_2-\hat n_3 \right)\,\nonumber\\
&&\times\,\, { B}^{\lambda_1 \lambda_2 \lambda_3}(f_1,\,f_2,\,\hat n_\star)\,,
 \label{builcor3}
\eea

\noindent
where in the last line we introduced the function ${ B}^{\lambda_{1,2,3}}$, the tensor bispectrum associated
to scenarios with stationary nG. Such bispectrum
is characterized by   flattened triangle shapes~\footnote{It is important to notice that  ${ B}^{\lambda_{1,2,3}}$ depends on a specific reference direction, that we denote with $\hat n_\star$: this
is due to the fact that tensor modes transform under spatial rotations, and the definition of polarization tensors depends on such  specific, selected direction. Our results for the overlap functions in the next Section 
then depend on the choice of $\hat n_\star$:
see also \cite{Bartolo:2018qqn} for a detailed discussion on this point.}  (see Fig \ref{fig:folded}).

For the rest of this work, we shall focus on stationary correlators of the form in eqs \eqref{builcor3a}, \eqref{builcor3}.  Folded non-Gaussianity is  known to arise in the scalar sector of specific
models of inflation, see e.g. \cite{Chen:2006nt,Holman:2007na,Bartolo:2010bj,Garcia-Saenz:2018vqf}. It would
be  interesting to investigate  models where the same shape of nG arise in the tensor sector, for example 
%
   in models with extra spin-2 degrees of freedom in an EFT approach to inflation
  (see e.g. \cite{Dimastrogiovanni:2018gkl,Goon:2018fyu}). We do not  pursue the problem of model building any
  further in this work, but we instead continue with characterizing the interesting properties of stationary graviton nG. 
   It is also worth noticing  that -- even if the background is isotropic -- the tensor bispectrum can distinguish among different chiralities, since its
amplitude depends on  the value of the chirality indexes $\lambda_i$. 

\bigskip

Substituting 
the 3-point function in Fourier space \eqref{builcor3} into eq \eqref{corrsta1}, we find the concise expression
\bea
\langle h_{a_1b_1} (t_1,\vec x_1) h_{a_2b_2} (t_2,\vec x_2)  h_{a_3 b_3} (t_3,\vec x_3)\rangle
&=&\sum_{\lambda_1 \lambda_2 \lambda_3}  \int_{-\infty}^\infty  d f_1  d f_2    \int d^2 \hat n
\, \times\,\e_{a_1b_1}^{(\lambda_1)}(\hat n)\,\e_{a_2b_2}^{(\lambda_2)}(\hat n)\,\,\e_{a_3b_3}^{(\lambda_3)}(\hat n)\,
\nonumber
\\
&\times& e^{2\pi\,i\,f_1\,\left(t_1-t_3\right)}\,
e^{2\pi\,i\,f_2\,\left(t_2-t_3\right)}\,e^{-2\pi\,i\,f_1\,\hat n\,\left(\,\vec x_1- \vec x_3\right)}\,e^{-2\pi\,i\,f_2\,
\hat n\,
\left(\vec x_2- \,\vec x_3\right)}
\nonumber
\\
&\times&
{ B}^{\lambda_1 \lambda_2 \lambda_3}(f_1,\,f_2,\,\hat n_\star)
\,,
\label{corrsta2}
\eea
that makes stationarity and isotropy particularly transparent.

\subsection{On the local observability of stationary graviton non-Gaussianity in a SGWB}

\begin{figure}[h!]
\begin{center}
\includegraphics[width = 0.35 \textwidth]{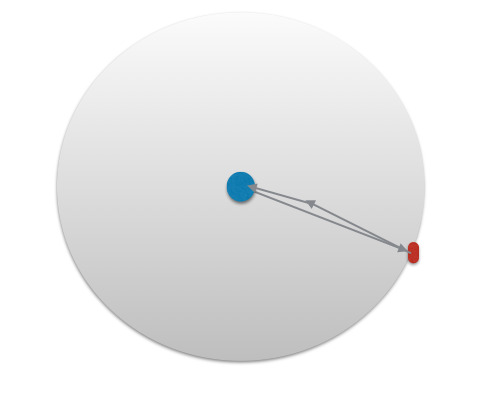}
 \caption{\it 
The structure of the 3-point function in eq \eqref{builcor3} requires that the three GWs entering in the correlator originate from
a common direction $\hat n$ in the sky. In the graphical representation above, we show with the red spot the common region of emission of three GWs (which
can be of cosmological origin); with the blue blob the region containing GW detectors (which can be of astrophysical size, as in the case of PTA experiments). The lines with arrows (that we intend as superimposed) indicate the GW common direction  $\hat n$.}
\label{fig:sphere}
\end{center} 
\end{figure}

We conclude this Section discussing some interesting properties of \eqref{corrsta2}. 
 Tensor 3-point   functions satisfying the stationarity condition \eqref{builcor3} do not necessarily suffer from  decorrelation  effects as  discussed in \cite{Bartolo:2018evs,Adshead:2009bz,Bartolo:2018rku,Allen:1996vm}. Such effects are associated with
   phase decorrelations among different 
 waves coming from several distinct  causally disconnected regions, a process that tends to `Gaussianize' the system for the central limit theorem. Interestingly, in our
 case, 
 the delta-function conditions on the wave-vectors given in  eq \eqref{builcor3}  (a consequence of stationarity)
 ensure us that GWs come from the same direction.  This is particularly clear from eq \eqref{corrsta2},
 where the angular integral is carried over the single direction of propagation of the waves (see also Fig \ref{fig:sphere}). 
   The work  \cite{Bartolo:2018evs} already pointed out that contributions to 3-point functions for which GW momenta are accurately aligned can avoid decorrelation effects. Our
   concept of stationary graviton nG  singles out the category of  tensor nG whose support is enhanced for such configurations, which are the only 
   ones that can be probed by measurements of 3-point functions of GW signals. 
  
 Besides this effect,  \cite{Bartolo:2018rku} shows that GWs, in their (possibly long)  way from source to detection, 
can 
  collect
  random   phases induced by long-wavelength energy fluctuations.
  These phases, physically  associated with a Shapiro time delay effect,
   influence the short-wavelength GW modes propagating over cosmological distances. Indeed, they 
     tend to suppress
 phase correlations of  
  initially  non-Gaussian fields, and to reduce the
 amplitude of connected
 $n$-point tensor correlation functions with $n\ge3$. In our case, correlators depend on time differences only. Hence  they are not sensitive
   to the entire time-travel of the wave from emission to detection, but only to the relatively short time-scale
   of the experiment. Although this is a more model dependent issue -- depending on how  long
   wavelength matter fluctuations influence tensor geodesics -- we can develop some semi-quantitative argument as follows.  
   The effect of long wavelength  modes can be expressed in terms of coordinate redefinitions \cite{Weinberg:2003sw}, which depend on time and on position. Following \cite{Bartolo:2018rku}, we focus on the
   effect of long wavelength curvature fluctuations, and describe their effect in terms of a shift of time coordinate.
  We express
the graviton mode in real space as
\be \label{sig_fourier1a}
h_{ab} (t,\vec x)\,=\,\sum_\lambda \int_{-\infty}^\infty  d f \int d^2 \hat n\,e^{-2\pi\,i\,f\,\hat n\,\vec x }
\,
e^{2\pi\,i\,f\,\left(t+Z(t,\,\hat n) \right)}\,
{\bf e}_{ab}^{(\lambda)}(\hat n)\,
 h_{\lambda}(f,\,\hat n)\,,
\ee
where the function $Z(t,\hat n)$ in the exponent  (depending on time and GW direction)  
characterizes the effect of the long mode.
 In taking the equal time 3-point function in
coordinate space using eq \eqref{sig_fourier1a}, and making use of stationary correlation properties  as in eq \eqref{builcor3},
the $\delta$-functions in the GW directions force all the arguments of the $Z$ functions
to be equal, and the $\delta$-functions in frequencies force them to cancel. The result  is not
dependent on time, nor on $Z$. Hence, long modes -- when  described as above -- do not influence the equal-time 3-point function. 
It would be interesting to formalize this argument more precisely, but such  analysis deserves more extensive work that
we leave to a future publication.

\section{Pulsar Timing Array overlap functions}
\label{sec_PTAoverlap}

We now investigate techniques to probe stationary tensor nG with pulsar timing arrays (from now on, PTA). 
 Precision measurements of time delays in pulsar periods can allow astronomers to extract interesting information
on the 
physics of the GW sector. Pulsar time delays can be due to a GW which deforming
the space-time by passing between the
pulsar and the earth;  to intrinsic pulsar period variations; or to  some unknown or less-known   noise  sources. 
By correlating measurements from distinct  pulsars,    noise can be reduced, and possible
GW signal revealed.  A correlation between  different time-delay measurements lead to 
the concept of {\it overlap function}, which quantifies the response  of a set of  GW detectors 
to 
GWs with a given frequency. Starting from the overlap function, it is then possible to estimate the signal-to-noise
ratio associated to dedicated GW observables aimed at characterise the non-Gaussian properties of a SGWB. 
  In this Section, after reviewing in Section \ref{sec2pt} well known results on 2-point overlap
functions for PTA observations, we pass to discuss 3-point functions specializing  to the  case
of stationary graviton non-Gaussianity. In particular:
\begin{enumerate}
\item In Section  \ref{sec3pt} we discuss overlap functions for pulsar timing arrays designed to
probe tensor non-Gaussianity with folded  shapes, corresponding to  stationary graviton non-Gaussianity
(see Section \ref{statio-nonG}). When correlating GW measurements from PTA experiments we expect the signals
to have comparable frequencies. Hence we probe flattened triangle shapes with comparable  side  lengths 
in momentum space, corresponding to the right panel of  Fig \ref{fig:folded}. \footnote{An interesting     study of PTA 3-point overlap functions, with the aim of 
to investigate tensor non-Gaussianity, 
 has been carried on in \cite{Tsuneto:2018tif}. But that work did not   specifically analyzed    flattened triangular shapes,
 that as we learned  are the physically relevant ones 
   in the context of stationary non-Gaussianity. \label{foot-comp}}
\item The tensor bispectrum can correlate also modes with different spins (e.g. tensors with scalars).
This  might lead to  interesting observables when investigating  theories of modified gravity with extra degrees
of freedom (as in scalar-tensor theories).  For the first time, in Section \ref{sec3ptScal} we compute mixed  3-point overlap functions
for GW experiments
 correlating tensor and scalar fluctuations, specialising to the case of PTA experiments.
\item A stationary  tensor 3-point function can also correlate GWs with very distinct frequencies, as long
as they satisfy the $\delta$-function constraints of eq \eqref{builcor3}: an example is the flattened triangle in
momentum space
of Fig \ref{fig:folded} (left panel) in which one of the frequencies
is much smaller than the others. This implies that triangle configurations can be probed by correlating different experiments
operating over different frequency ranges.  For the first time, in Section \ref{corrGB} 
we build overlap functions correlating
distinct experiments: PTA  (detecting SGWBs at frequencies
of order $f_{\rm PTA}\,\sim\,10^{-9} - 10^{-7}$ Hz) and ground based detectors (operating at frequencies
of $f_{\rm GB}\,\sim\,10^{0} - 10^{3}$ Hz). 
\end{enumerate}

\bigskip
Following the review in \cite{Maggiore:2018sht},
 we define the total time-delay output $s_\alpha$ measured by a GW experiment based on
a PTA system
  as sum of a GW signal $\sigma_\alpha$, and the noise $n_a$. We assume that the noise
is uncorrelated with the GW signal, and both have average zero.  
For any GW propagating in the direction $\hat n$, we define the signal detected by the PTA in terms of the 
  relative time delay induced by the GW 
on the pulsar period 
\be
\sigma_\alpha(t)\,\equiv\,\frac{\Delta T_\alpha}{T_\alpha}\,=\,\frac{x_\alpha^i\,x_\alpha^j}{2(1+\hat n \cdot \hat x_\alpha)}\,\left[h_{ij}(t,\,\vec x=0)-h_{ij}(t-\tau_\alpha,\,\vec x_\alpha)
\right]\,,
\ee
where the Earth is located at position $\vec x\,=\,0$, while the pulsar $\alpha$ is located at  position $\vec x\,=\,\vec x_\alpha$. $\tau_\alpha$ is the light travel time from the pulsar to the Earth and $\hat x_a$ the unit vector between the Earth and the pulsar position. 
Expressing this quantity in Fourier space, we find
\be\label{sigFou}
\sigma_\alpha(t)\,=\,
\sum_\lambda \int_{-\infty}^\infty  d f \int d^2 \hat n\
F^{(\lambda)}_\alpha(\hat n)
\,
e^{2\pi\,i\,f\,t}\,
 h_{\lambda}(f,\,\hat n)
 \,\left( 1-e^{-2 \pi i f \tau_\alpha}  \, e^{-2\pi\,i\,f\,\hat n\,\vec x_\alpha } \right)\,,
\ee
where we introduce the detector tensor  
\be \label{dtPTA}
F_\alpha^{(\lambda)}(\hat n)\,=\,\frac{x_\alpha^i\,x_\alpha^j\,{\bf e}_{ij}^{(\lambda)} (\hat n)}{2(1+\hat n \cdot \hat  x_\alpha)}\,\,,
\ee
which depends only on the GW direction, but not on the GW frequency $f$.

\subsection{Two-point overlap functions}\label{sec2pt}

The simplest possibility to consider   is the  2-point correlation function. We review this well known
case here, before discussing new results for the 3-point overlap function for stationary nG.  
The 2-point function for the GW modes is given in eq \eqref{builcor3a}. 
The  equal-time
2-point correlation function for the PTA time-delay signal reads
\bea
\langle \sigma_\alpha (t) \sigma_\beta (t) \rangle
&=&
\sum_{\lambda_1 \lambda_2} 
\int_{-\infty}^\infty  d f_1 d f_2  \int d^2 \hat n_1 d^2 \hat n_2\
F^{(\lambda_1)}_\alpha(\hat n_1) F^{(\lambda_2)}_\beta(\hat n_2)
\,
e^{2\pi\,i\,\left(f_1+f_2 \right)\,t}\,
 P( f_1)
  \nonumber
 \\
 &&\times \,  \,\delta^{(2)}(\hat n_1-\hat n_2)\, \delta(f_1+f_2)\,\delta_{\lambda_1 \lambda_2}\,
\left( 1-e^{-2 \pi i f_1 \tau_\alpha\left(1+ \hat n_1  \,\hat n_\alpha \right)}  \right)
  \,\left( 1-e^{-2 \pi i f_2 \tau_\beta\left(1+ \hat n_2  \,\hat n_\beta \right)}   \right)\,.
  \nonumber
 \\
\eea
For pulsars at typical distances of $10^3$ parsec, taking
into account the  frequency range probed by PTA, one finds that the quantity $ f \tau_\alpha\sim {\cal O}(10^2)$. This implies that  contributions containing  exponentials in the pulsar terms, in the second line of
the previous equation,  are rapidly oscillating
functions that are  averaged out in the {integral over directions $\hat n$}. 
 Hence
 we can neglect these terms and substitute the
second line with a unit factor.

\begin{figure}[H]
\begin{center}
\includegraphics[width = 0.3 \textwidth]{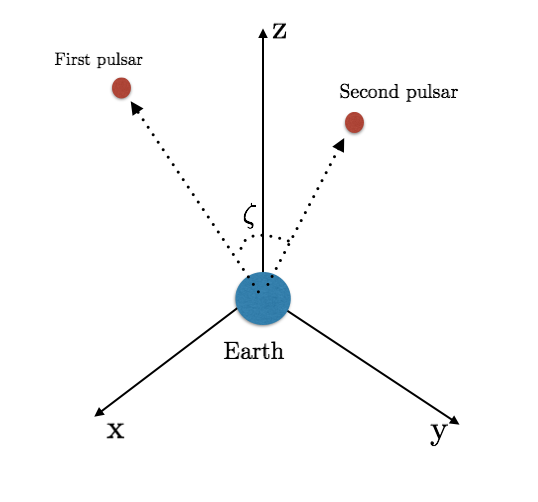}
 \caption{\it 
Representation of a system of two pulsars, and the earth. We denote with $\zeta$ the angle between the unit vectors from the earth towards each of the pulsars.}
\label{fig:art1}
\end{center} 
\end{figure}

\begin{figure}[H]
\begin{center}
\includegraphics[width = 0.4 \textwidth]{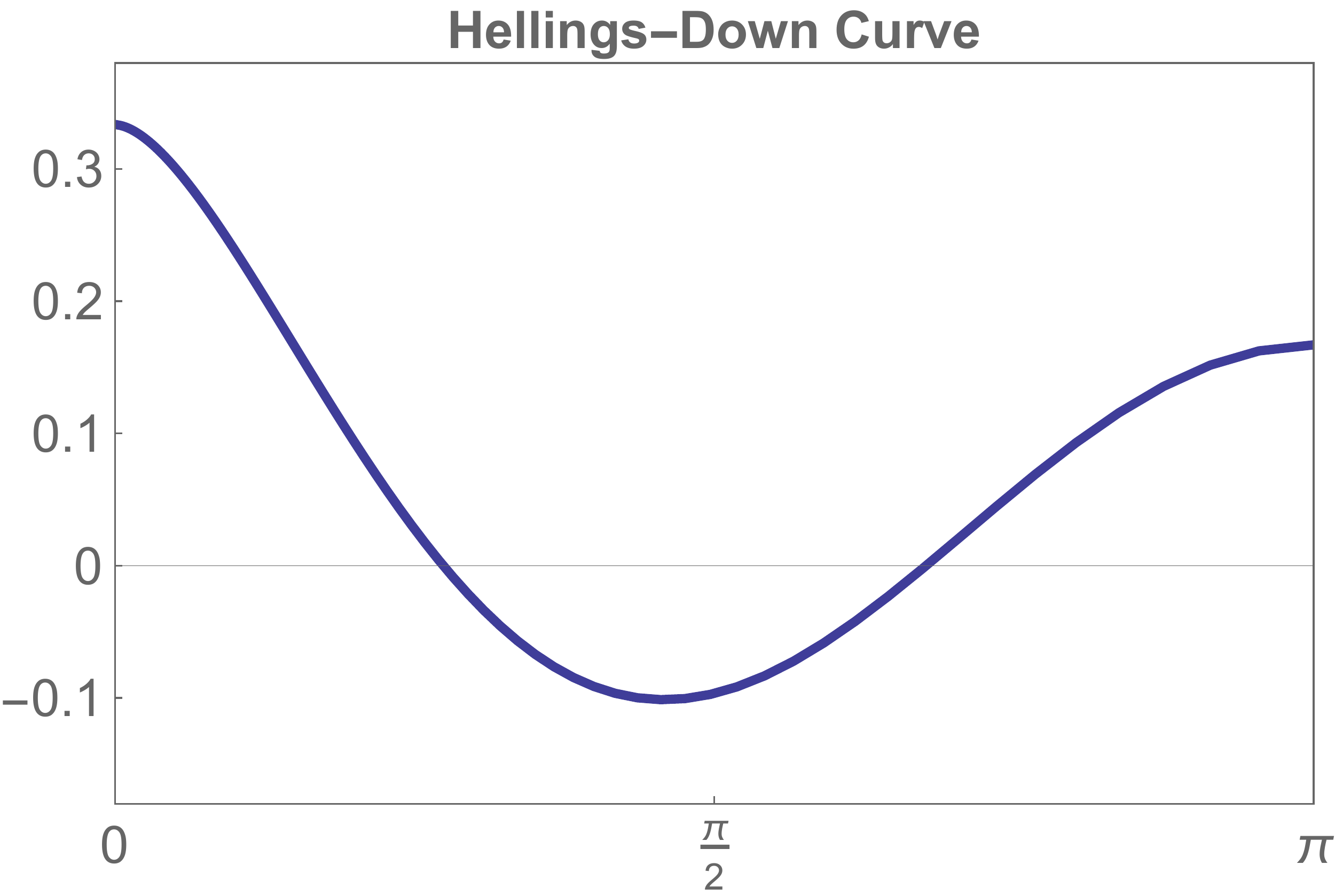}
 \caption{\it 
The 2-point overlap function for the
sum of the  spin-2 polarizations  
$ {\cal R}_{ab}(\zeta)\,=\,\sum_\lambda {\cal R}^{(\lambda)}_{ab}$ (the so-called
Hellings-Down curve \cite{Hellings:1983fr}). 
The $x$-axis contain
the angle $\zeta$ defined in eq \eqref{defze1}. The $y$-axis the corresponding value of the overlap function,
see eq \eqref{defRlam}.}
\label{fig:HDcl}
\end{center} 
\end{figure}

Under this approximation, making use of the $\delta$ functions, we can assemble the angular integral into 
 a response function, and 
express the previous quantity as
\bea
\langle \sigma_\alpha (t) \sigma_b (t) \rangle&=& 2\pi\,\sum_\lambda \int d f\, {\cal R}^{(\lambda)}_{\alpha \beta}
\,P(f)\,,
\eea
where the PTA 2pt response function is given by an angular integration, leading to  the so-called 
Hellings-Down overlap function \cite{Hellings:1983fr} (see also \cite{Anholm:2008wy,Mingarelli:2016zug,Maggiore:2018sht}):
\bea \label{defRlam}
 {\cal R}^{(\lambda)}_{\alpha \beta}(\zeta)&=&\int \frac{d^2 \hat n}{2 \pi}\,F^{(\lambda)}_\alpha(\hat n) F^{(\lambda)}_\beta(\hat n)\,,
 \\
 &=&\frac16-\frac{\left(1-\cos \zeta \right)}{24}\left(1 -6\,\ln \left( \frac{1-\cos \zeta }{2}\right)\right)\,,
\eea
 where
 \be \label{defze1}
\cos \zeta_{}\,\equiv\, \hat x_\alpha\,\cdot\,\hat x_\beta\,.
\ee 
It is important to stress that the  Hellings-Down  function $ {\cal R}^{(\lambda)}_{\alpha \beta}$ depends only on angle $\zeta$ between pulsar directions,  and not on the GW frequency.  In Fig \ref{fig:HDcl} we  represent the profile
of the sum of overlap functions over polarization indexes,  
$ {\cal R}_{\alpha \beta}(\zeta)\,=\,\sum_\lambda {\cal R}^{(\lambda)}_{\alpha \beta}$.


\subsection{Three-point overlap functions}\label{sec3pt}

We now analyze the 3-point overlap function for flattend tensor bispectra satisfying 
a stationarity condition, as described in Section \ref{statio-nonG}. Such
shapes of tensor bispectra were not specifically investigated  in \cite{Tsuneto:2018tif}, hence our results
are new. In correlating
signals from different pulsars, we assume
that the frequencies of the GWs are comparable, hence the 
flattened tensor bispectra in momentum space have a shape with sides
 of similar  size, see the right panel of
 Fig \ref{fig:folded}.
 
 The tensor 3-point function in
Fourier space is given by eq \eqref{builcor3}. 
The PTA signal equal-time
3pt  function then results, using the notation of the previous subsection, is 
\bea
\langle \sigma_\alpha (t) \sigma_\beta (t)
\sigma_\gamma (t) 
 \rangle
&=&
\sum_{\lambda_1 \lambda_2 \lambda_3} 
\int_{-\infty}^\infty  d f_1 d f_2  \int d^2 \hat n
\, F^{(\lambda_1)}_\alpha(\hat n) F^{(\lambda_2)}_\beta(\hat n)
F^{(\lambda_3)}_\gamma(\hat n)
\,
 { B}^{\lambda_1 \lambda_2 \lambda_3}( f_1, f_2, \hat n_\star)\,.\nonumber
 \\
 \eea
 We now conveniently  collect the angular integration in the definition of the 3-point overlap
 function ${\cal R}_{ \alpha \beta \gamma}^{\lambda_1 \lambda_2 \lambda_3} (\hat n_\star)$: 
 \bea
\langle \sigma_\alpha (t) \sigma_\beta (t)
\sigma_\gamma (t) 
 \rangle
&=&
\sum_{\lambda_1 \lambda_2 \lambda_3} 
2 \pi\,\int_{-\infty}^\infty  d f_1 d f_2 \,
{\cal R}_{ \alpha \beta \gamma}^{\lambda_1 \lambda_2 \lambda_3}( \hat n_\star)
\,
 { B}^{\lambda_1 \lambda_2 \lambda_3}( f_1, f_2, \hat n_\star)\,.
\eea
To perform such angular integrations, we introduce a rotation matrix $M\left[ \theta ,\, \phi   \right]$:
\be
M\left[ \theta ,\, \phi   \right] = 
\left( \begin{array}{ccc} 
\sin \theta \cos \phi  & \cos \theta  \cos \phi  & - \sin \phi   \\ 
\sin \theta  \sin \phi  & \cos \theta  \sin \phi  & \cos \phi    \\  
\cos \theta  & - \sin \theta  & 0 
\end{array} \right)  \,.
\label{rotmafc}
\ee
We rotate over the reference direction $n_\star$:  
$\hat n\,=\,M\left[ \theta ,\, \phi   \right]\,\hat n_\star$, so to write
\be\label{3ptOv}
{\cal R}^{\lambda_1 \lambda_2 \lambda_3}_{\alpha\beta\gamma}( \hat n_\star)
\,=\,\frac{1}{2\pi}\int_0^{2\pi}\,d \phi\,\,\int_0^\pi\,\sin{\theta}\,d \theta\, \left[
F^{(\lambda_1)}_\alpha(M\,\hat n_\star) F^{(\lambda_2)}_\beta(M\,\hat n_\star)
F^{(\lambda_3)}_\gamma(M\,\hat n_\star)+\left( \lambda_i\to-\lambda_i\right)\right]\,.
\ee
Notice that the PTA 3-point overlap function
 depends on the chirality  of the GW,  the relative position of pulsars,  and  the reference
direction $\hat n_\star$. From now on, in this work we select
$\hat n_\star$ to point along the $\hat x$ axis:
\be
\hat n_\star\,=\,\left(1,\,0,\,0 \right)\,.
\ee
  On the other hand, the overlap function does {\it not} depend 
on the frequency since (for
the same arguments discussed in Section \ref{sec2pt}) we can safely neglect
the earth terms.   In what follows, we do not consider scenarios
including parity violation, and we sum over opposite
chiralities. In other words, indicating $R\,=\,+$ and $L\,=\,-$, we compute and plot 
the sum ${\cal R}^{\lambda_1 \lambda_2 \lambda_3}_{\alpha\beta\gamma}+{\cal R}^{-\lambda_1\,-\lambda_2 \,-\lambda_3}_{\alpha\beta\gamma}$, as done in Section \ref{sec2pt} for the 2-point overlap function. 

Evaluating the precise angular structure of  ${\cal R}^{\lambda_1 \lambda_2 \lambda_3}_{\alpha\beta\gamma}( \hat n_\star)$ is essential for estimating the optimal signal-to-noise ratio
of an experiment to measure a stationary tensor  bispectrum, as we shall learn in Section \ref{sec-optimal}. In the
next two subsections, we evaluate the 3-point overlap functions in two different  configurations.

\subsubsection{3-point overlap function: two signals from the same   pulsar}\label{secsame}

\begin{figure}[H]
\begin{center}
\includegraphics[width = 0.38 \textwidth]{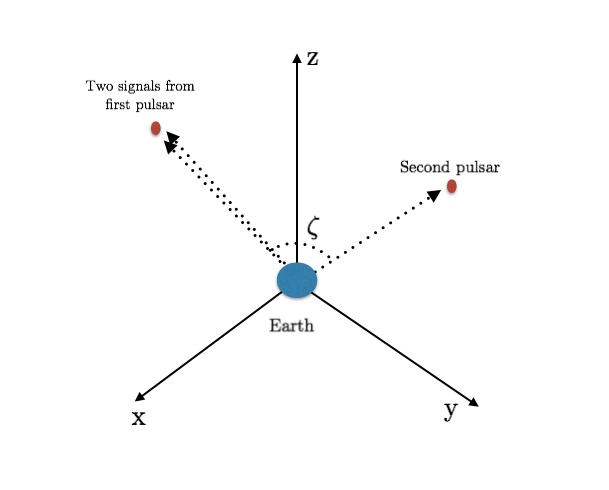}
 \caption{\it 
Representation of a  configuration where we correlate two signals measured at the same pulsar $\alpha$ with a  signal for the distinct  pulsar $\beta$.}
\label{fig:art4}
\end{center} 
\end{figure}

In this subsection we compute the overlap function for a case where we correlate two time-delay signals
from the same pulsar $\alpha$, with a third signal  from a separate pulsar $\beta$. This 
case will be important for the discussion  in Section \ref{sec-optimal}. 
 We consider
the limit of  equal time 3-point  correlation
\be
\left\langle \sigma_\alpha \left( t \right) \sigma_\alpha \left( t \right) \sigma_\beta \left( t \right) \right\rangle\,,
\ee
and we represent in Fig \ref{fig:art4} for a graphical representation of the geometry of the system.

\begin{figure}[H]
\begin{center}
\includegraphics[width = 0.45 \textwidth]{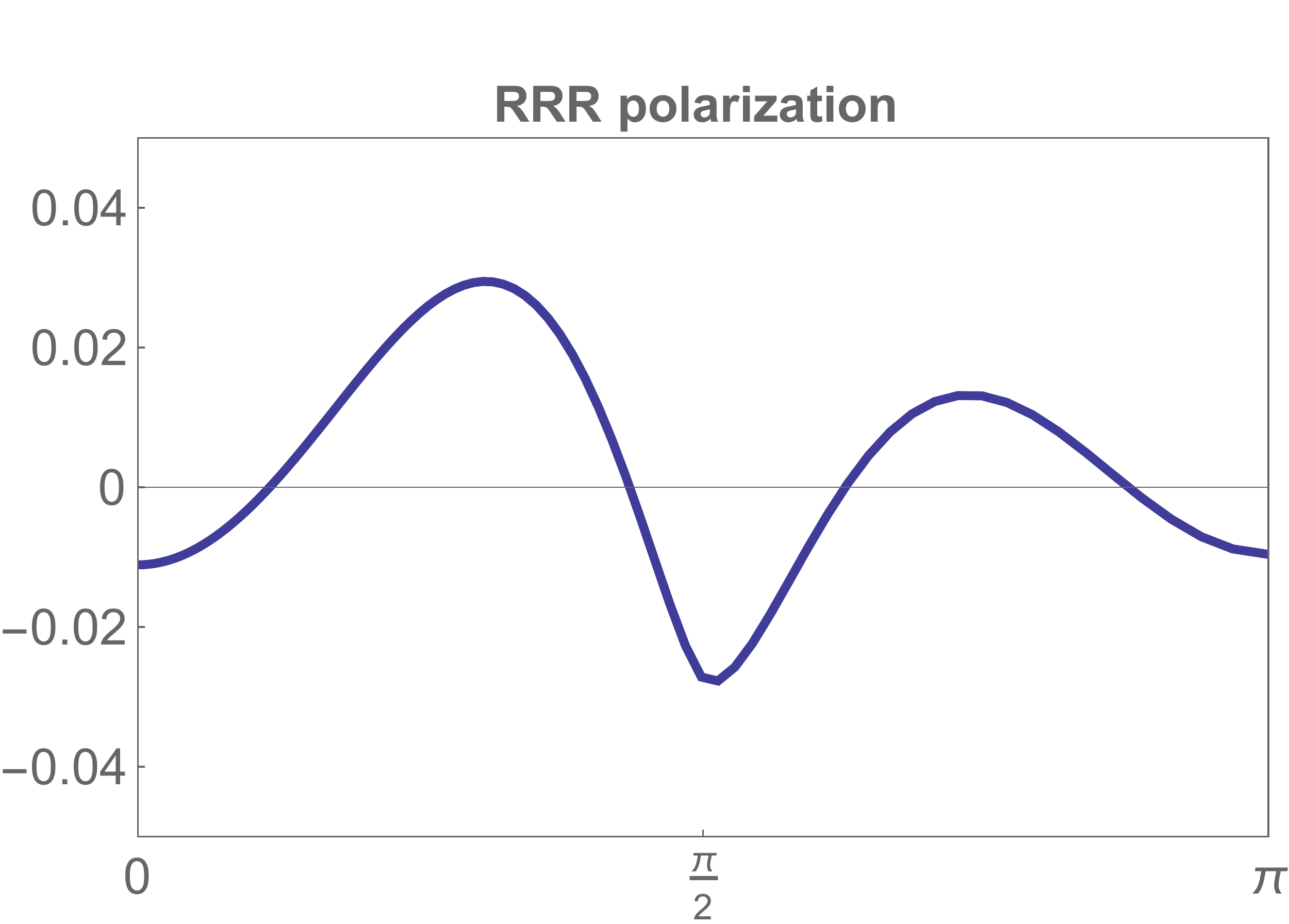}
\includegraphics[width = 0.45 \textwidth]{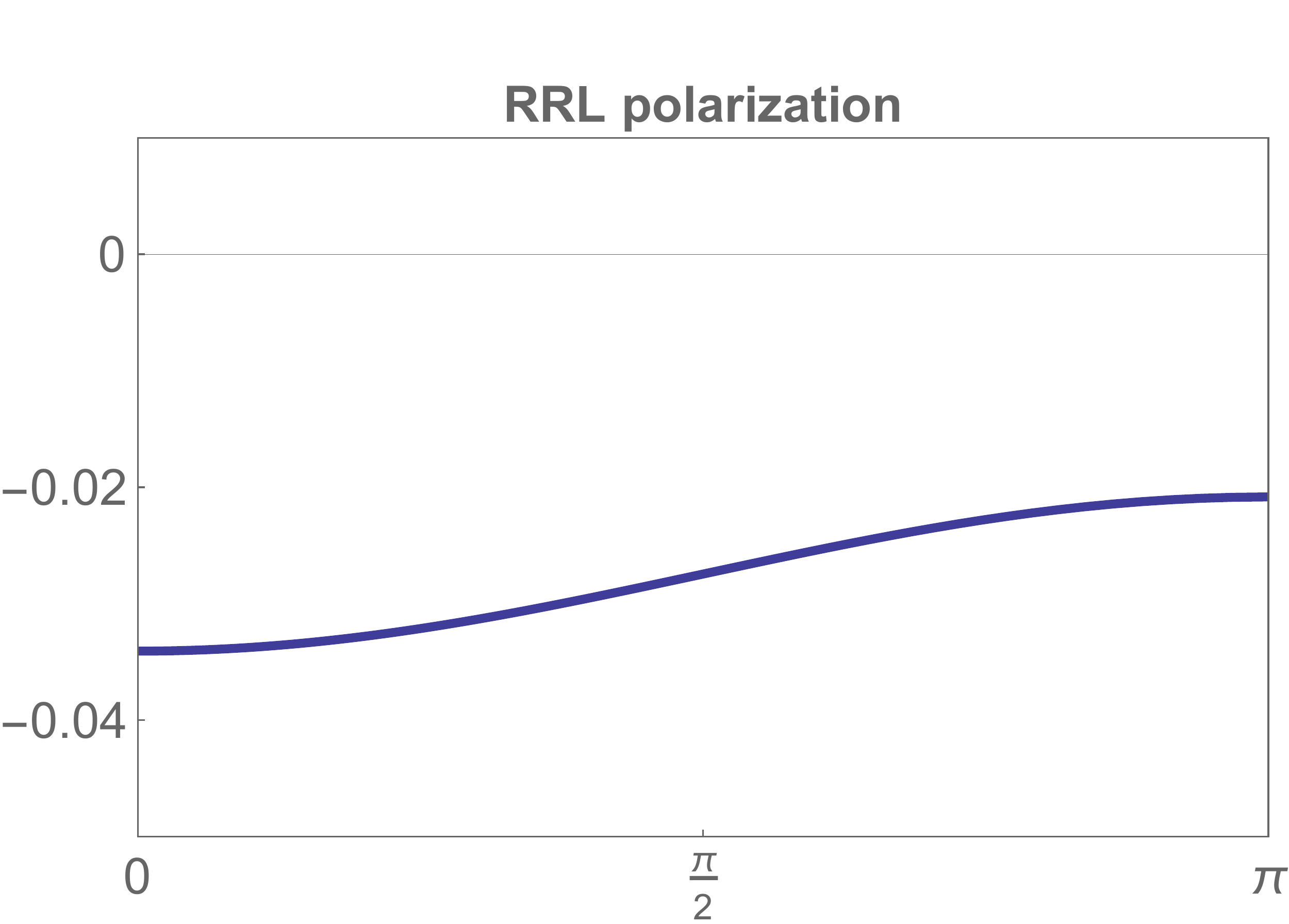}
 \caption{\it 3-point overlap  functions associated with stationary, flattened bispectra,  for correlating  two signals from the same pulsar  with a signal from another
 pulsar (see Fig \ref{fig:art4}). In the x-axis we vary the angle $\zeta$  between 
 the
 unit vectors from the earth towards pulsars $\alpha$ and $\beta$. In the y-axis we represent the magnitude of
 the associated 3-point function, which depends on the polarization of the GWs.
 %
 }
\label{3dpl0A} 
\end{center} 
\end{figure}

\noindent
In this case we find that the 3-point  overlap function depends only on the angle $\zeta$ between the vectors from the earth towards pulsars $\alpha$ and $\beta$. The general formula  to compute such function
is given by eq \eqref{3ptOv}, with $\gamma=\beta$.
 The  
 response function $ {\cal R}_{\alpha\beta\beta}^{\lambda_1 \lambda_2\lambda_3}$ depends on chirality, see Fig \ref{3dpl0A}. Its magnitude is around few percent, depending on the value of $\zeta$. We find ${\cal R}_{\alpha \beta \beta}^{RRR}\,=\, {\cal R}_{\alpha \beta \beta}^{RLR}$. Notice that these  results are quite different from the Hellings-Down profile reviewed   in Section \ref{sec2pt} for 
 the 2-point function, 
  in respect of size and angular profile.

\subsubsection{3-point overlap function: signals from  from three distinct  pulsars}\label{sub-ortho}

\begin{figure}[H]
\begin{center}
\includegraphics[width = 0.28 \textwidth]{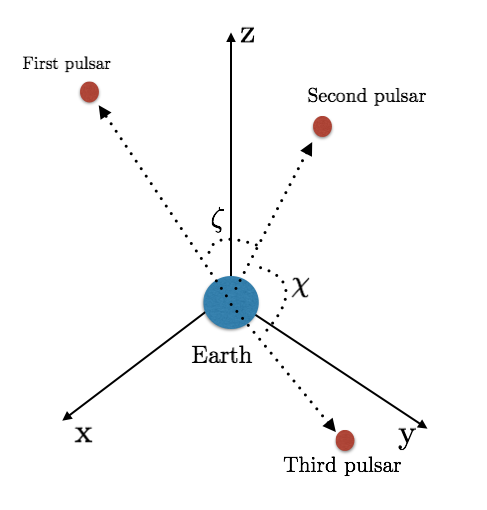}
 \caption{\it 
A configuration with three pulsars lying on orthogonal planes. In this representative figure, $\zeta$ denotes the angle beween the unit vectors from the earth towards pulsar 1 and 2. $\chi$ denotes the angle beween the unit vectors from the earth towards pulsar 2 and 3.}
\label{fig:art2a}
\end{center} 
\end{figure}

When we correlate time-delay signals from three different pulsars, the result depends 
in a more complex way on the geometry of the system. For definiteness,  here  we focus
our attention on a system where the three pulsars lie  
 on orthogonal planes, $(x,\,y)$, $(x,\,z)$, $(y,\,z)$: see Fig \ref{fig:art2a}.

\begin{figure}[H]
\begin{center}
\includegraphics[width = 0.3 \textwidth]{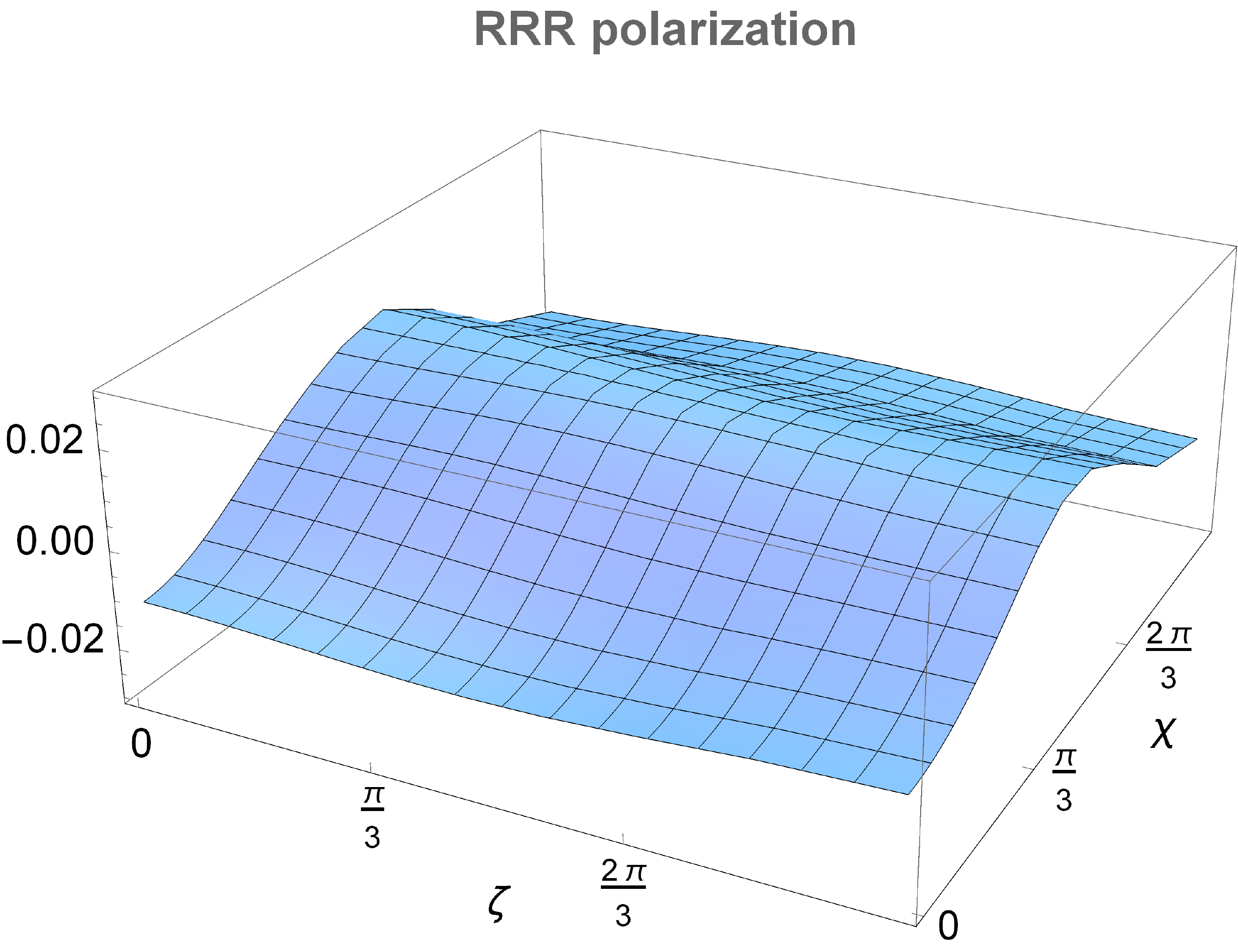}
\includegraphics[width = 0.3 \textwidth]{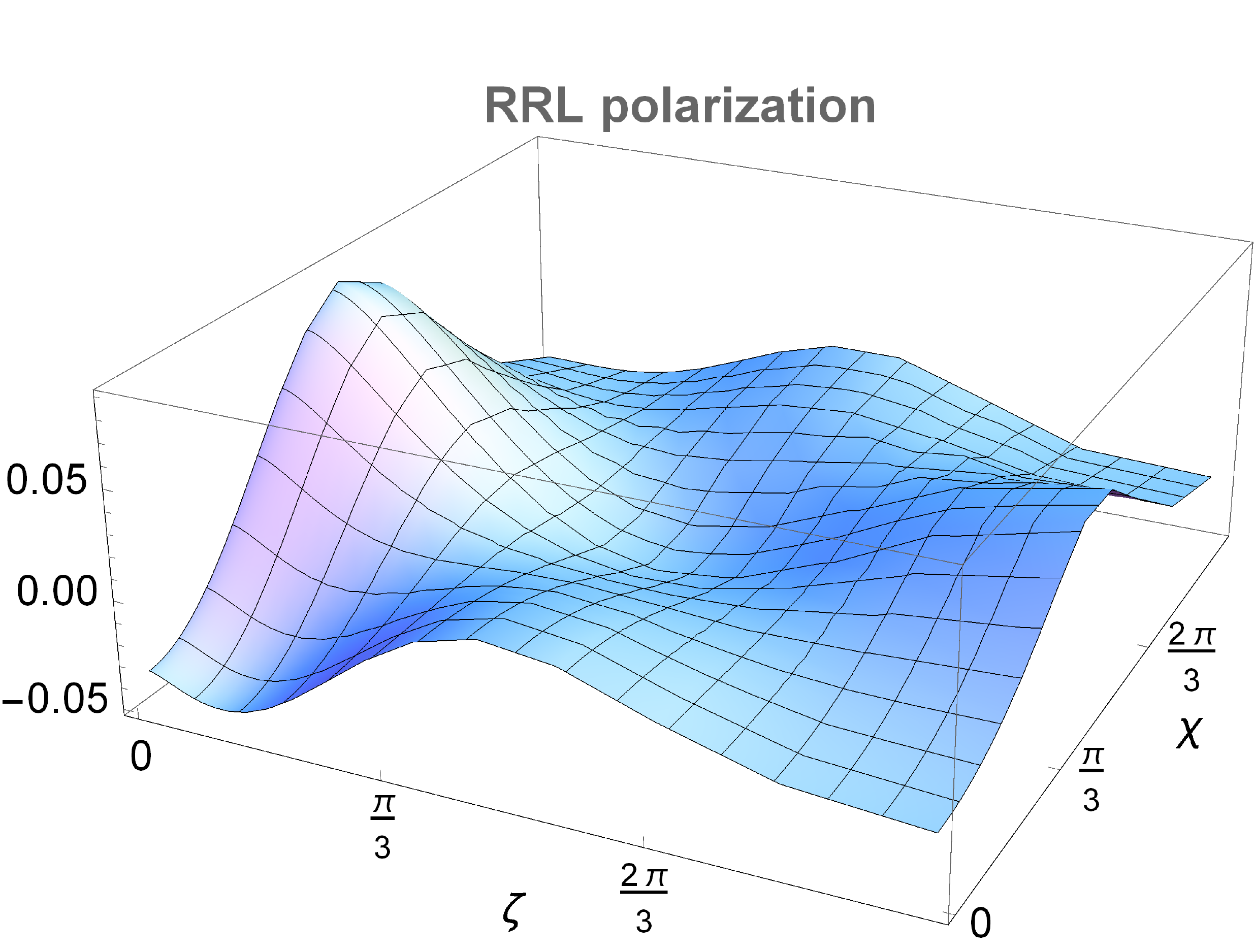}
\includegraphics[width = 0.3 \textwidth]{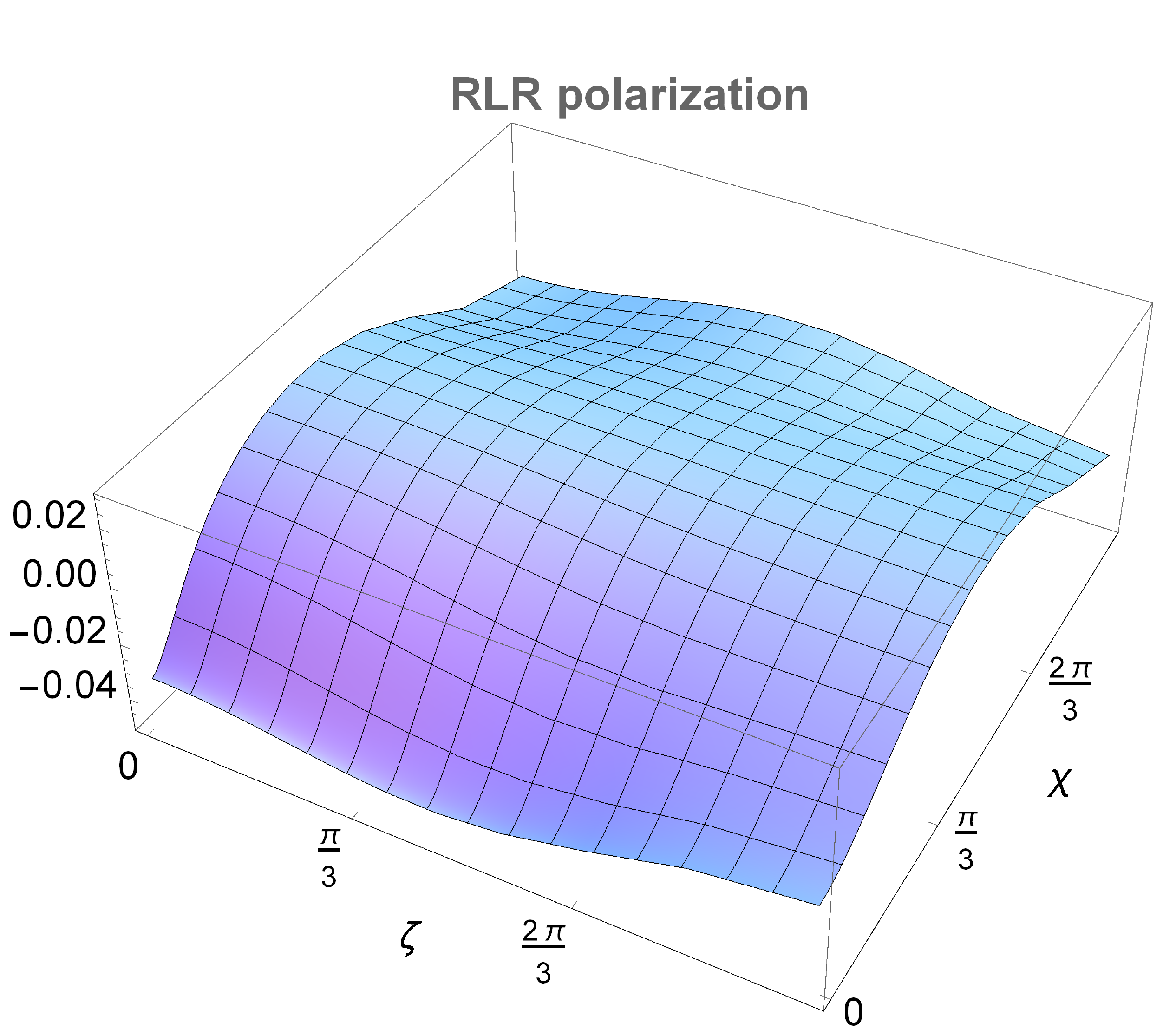}
 \caption{\it 
  Three-point overlap  functions associated with stationary, flattened bispectra
   We locate the pulsars in three different orthogonal planes (say (xy), (yz), (xz)) and vary the angle $\zeta$  between 
 the
 unit vectors from the earth towards pulsar 1 and 2,
 and $\chi$  the unit vectors from the earth towards pulsar 1 and 3. The z-axis represents
 the magnitude of the corresponding overlap function. 
}
\label{3dpl1} 
\end{center} 
\end{figure}

\noindent
In this case, the 3-point 
 overlap function depends on two angles $\zeta$ and $\chi$ between the vectors from earth towards the pulsars.
 We represent our results for the overlap functions in 
 Fig \ref{3dpl1}, evaluated using the general formula \eqref{3ptOv}. Also in this case the typical magnitude of the overlap function is of order of a few percent. On the other hand, the plots in Fig   \ref{3dpl1} also present
 peaks and valleys where the magnitude of the overlap function can increase by 
  a factor of 
 order one  
 with respect to the 2-dimensional case studied in section \ref{secsame}.

\subsection{Three-point overlap functions: correlating tensor  and scalar polarizations}\label{sec3ptScal}


GW experiments can be sensitive also to vector and scalar polarizations of GWs,  motivated by
 theories that modify General Relativity: see the interesting early works \cite{Eardley:1973br,Eardley:1974nw}
  that first explored this possibility, and \cite{Yunes:2013dva,Chamberlin:2011ev,Cornish:2017oic}
   for analysis that specifically cover
 PTA experiments.  GW 2-point functions do not correlate
fields of different spin (e.g. scalar and tensors) around an isotropic background; instead, correlations
among different spins are possible at the level of  3-point
functions. In this subsection, for the first time we investigate   3-point overlap functions describing the correlation
of scalar GW polarizations with tensor (or scalar) polarizations. Our results are  model independent, and we 
do not refer to specific scenarios. But, for the same reasons discussed in the previous sections, we focus on
stationary non-Gaussianity, associated with time-translation-invariant 3-point functions (that is, bispectra
with folded shapes in momentum space). 
 Our  conventions for polarization
tensors describing the transverse scalar  `breathing mode' GW polarization are standard and listed  in Appendix \ref{app:conventions}. 

\begin{figure}[H]
\begin{center}
\includegraphics[width = 0.3 \textwidth]{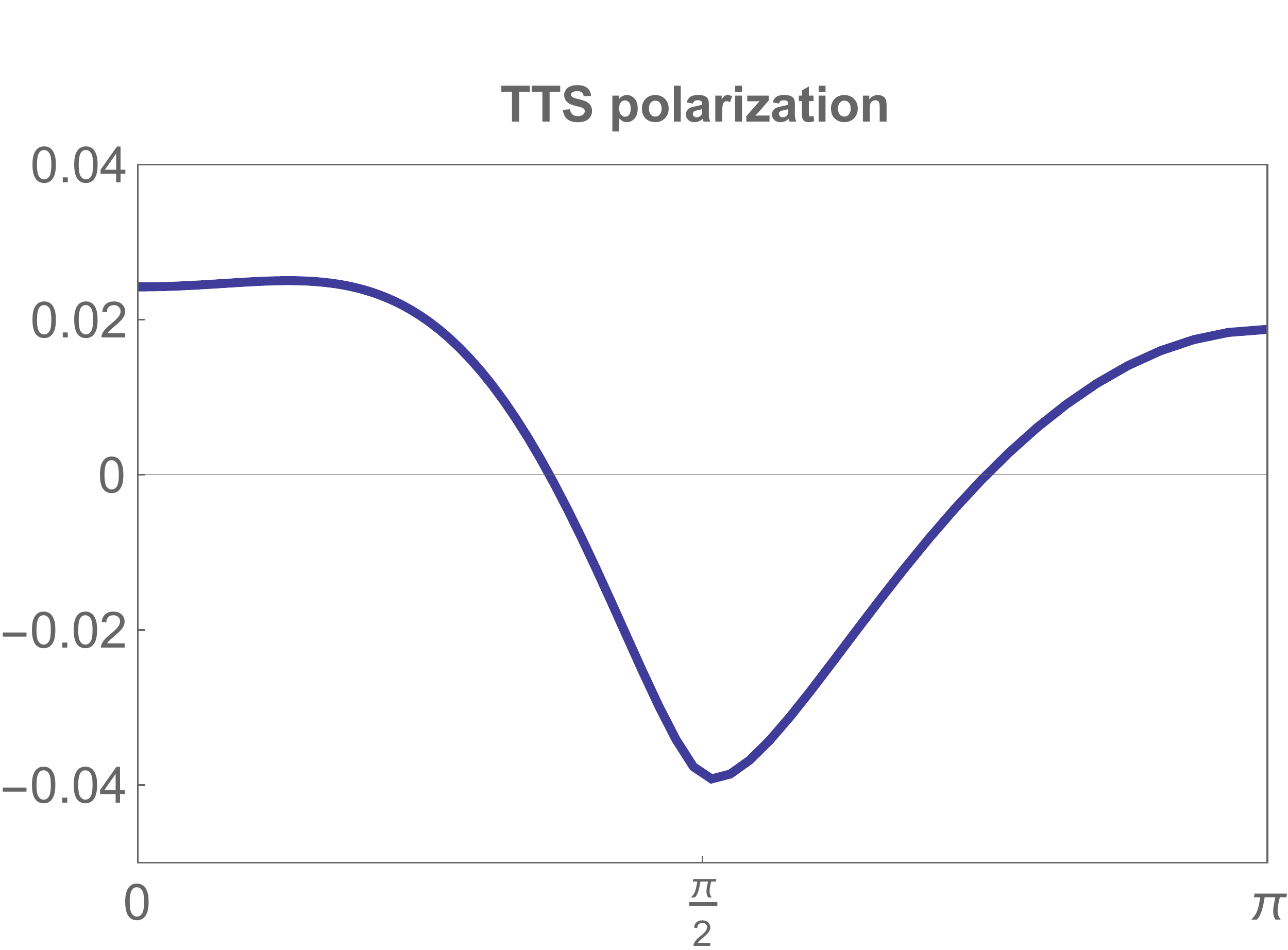}
\includegraphics[width = 0.3 \textwidth]{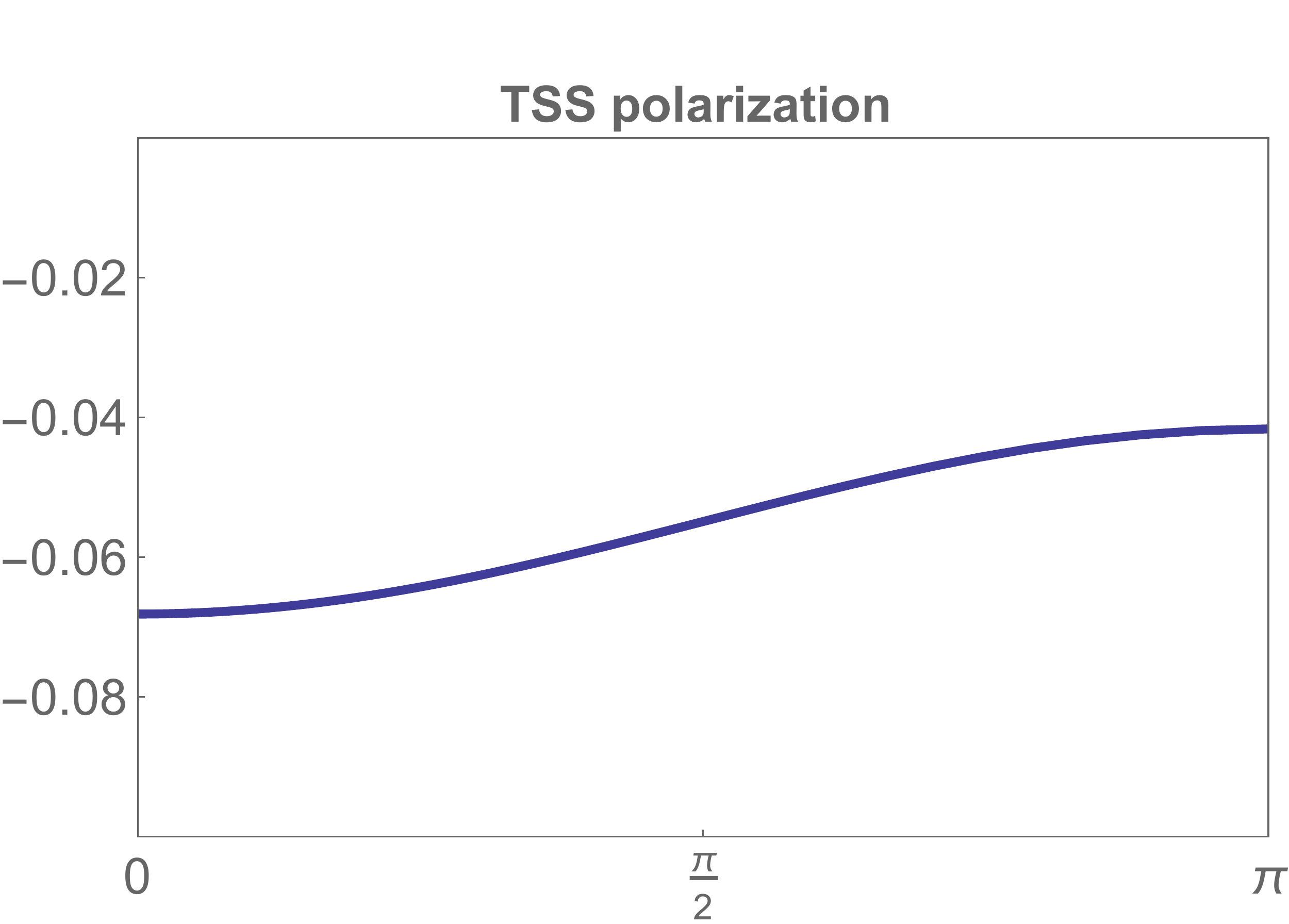}
\includegraphics[width = 0.3 \textwidth]{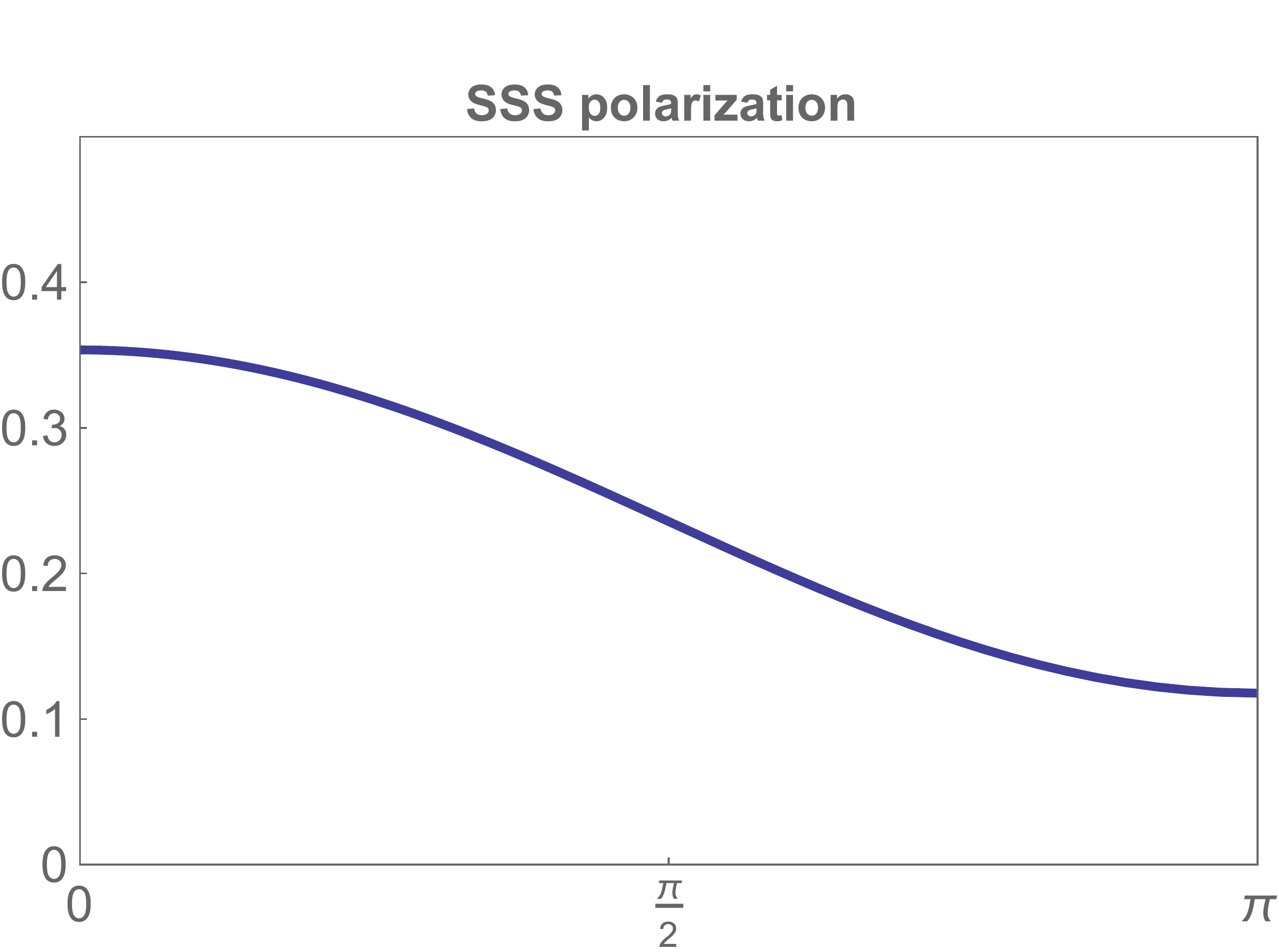}
 \caption{\it Three-point overlap  functions  correlating  two signals from the same pulsar $\alpha$  with a signal from second
 pulsar $\beta$. We vary the angle $\zeta$  between 
 the
 unit vectors from the earth towards pulsar a and b. We correlate tensor {\bf T}  polarization  with scalar {\bf S} polarization. 
}
\label{3dpl0s} 
\end{center} 
\end{figure}

We first consider correlation of two signals from the same pulsar $\alpha$ with
a signal from a distinct pulsar $\beta$, as in subsection \ref{secsame}. Figure \ref{3dpl0s} shows our result
for the overlap function relative to a stationary bispectrum of folded shape, with correlate
scalar and tensor modes. Interestingly, the magnitude of  the overlap function in the presence
of scalar excitations is larger than in the case of spin-2 modes only, and does not depend on the chirality (L/R) of tensors entering the correlators. 
 As we shall see, this fact has some consequences in the computation of the signal-to-noise
ratio.

\begin{figure}[H]
\begin{center}
\includegraphics[width = 0.3 \textwidth]{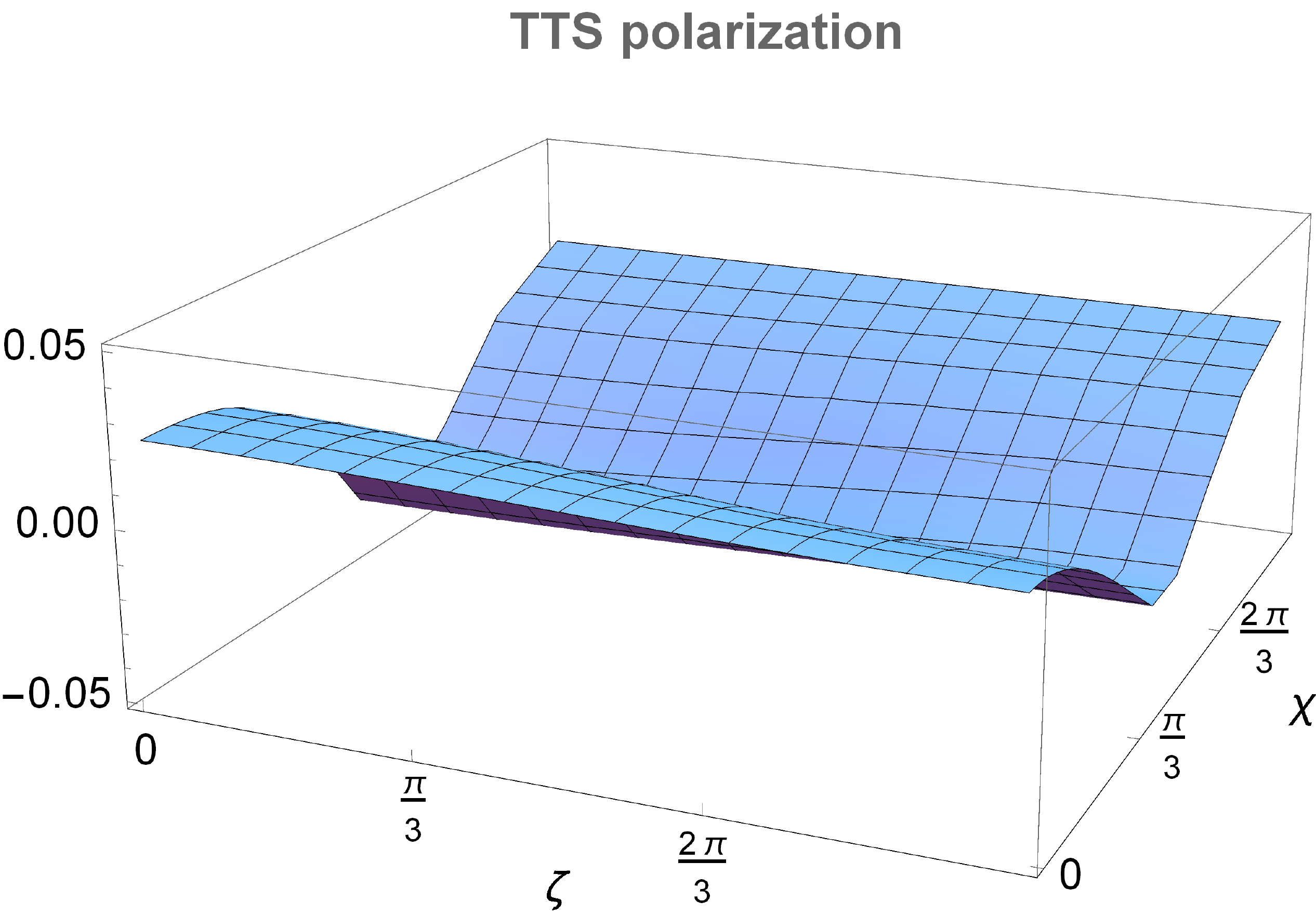}
\includegraphics[width = 0.3 \textwidth]{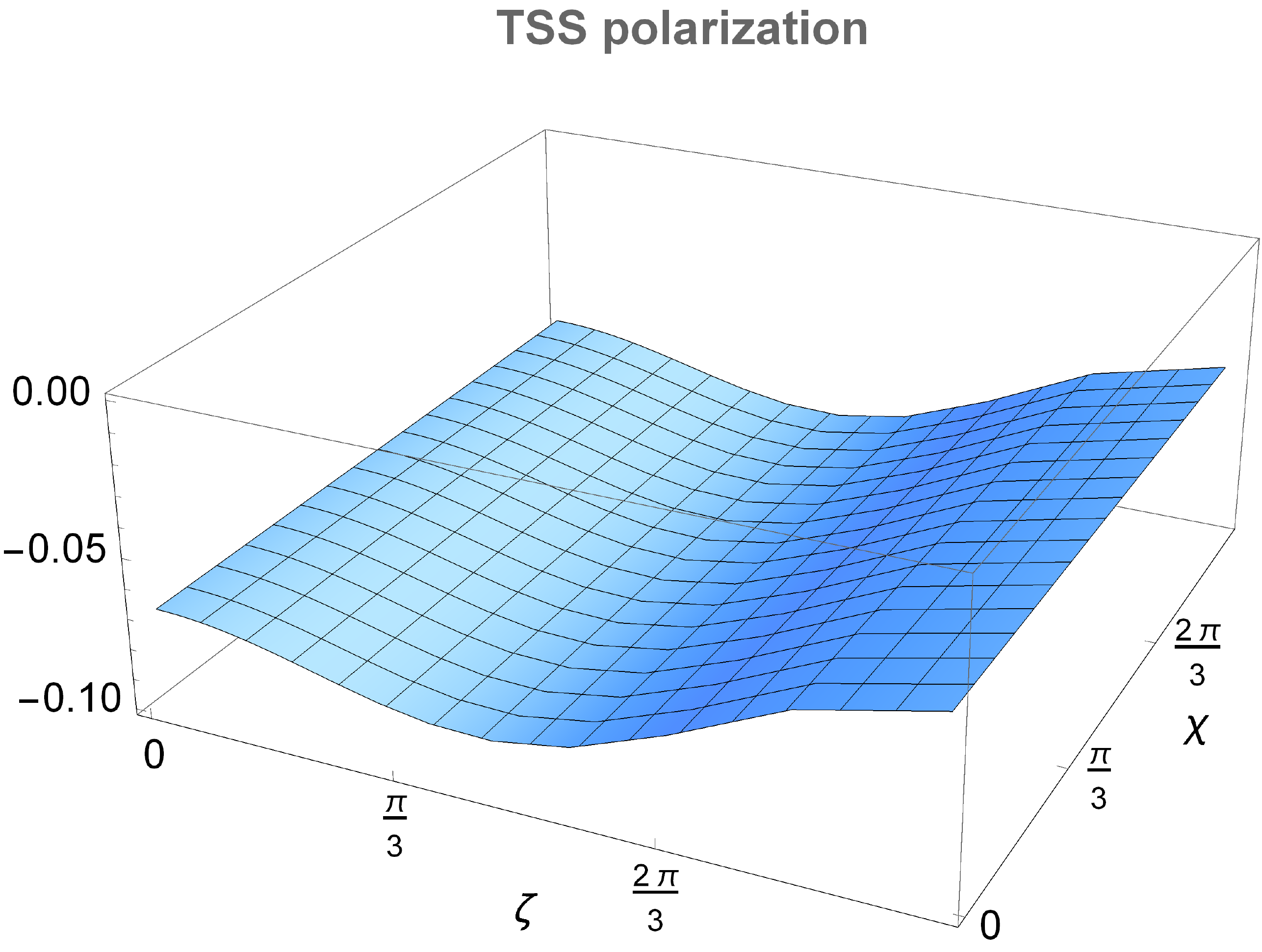}
\includegraphics[width = 0.3 \textwidth]{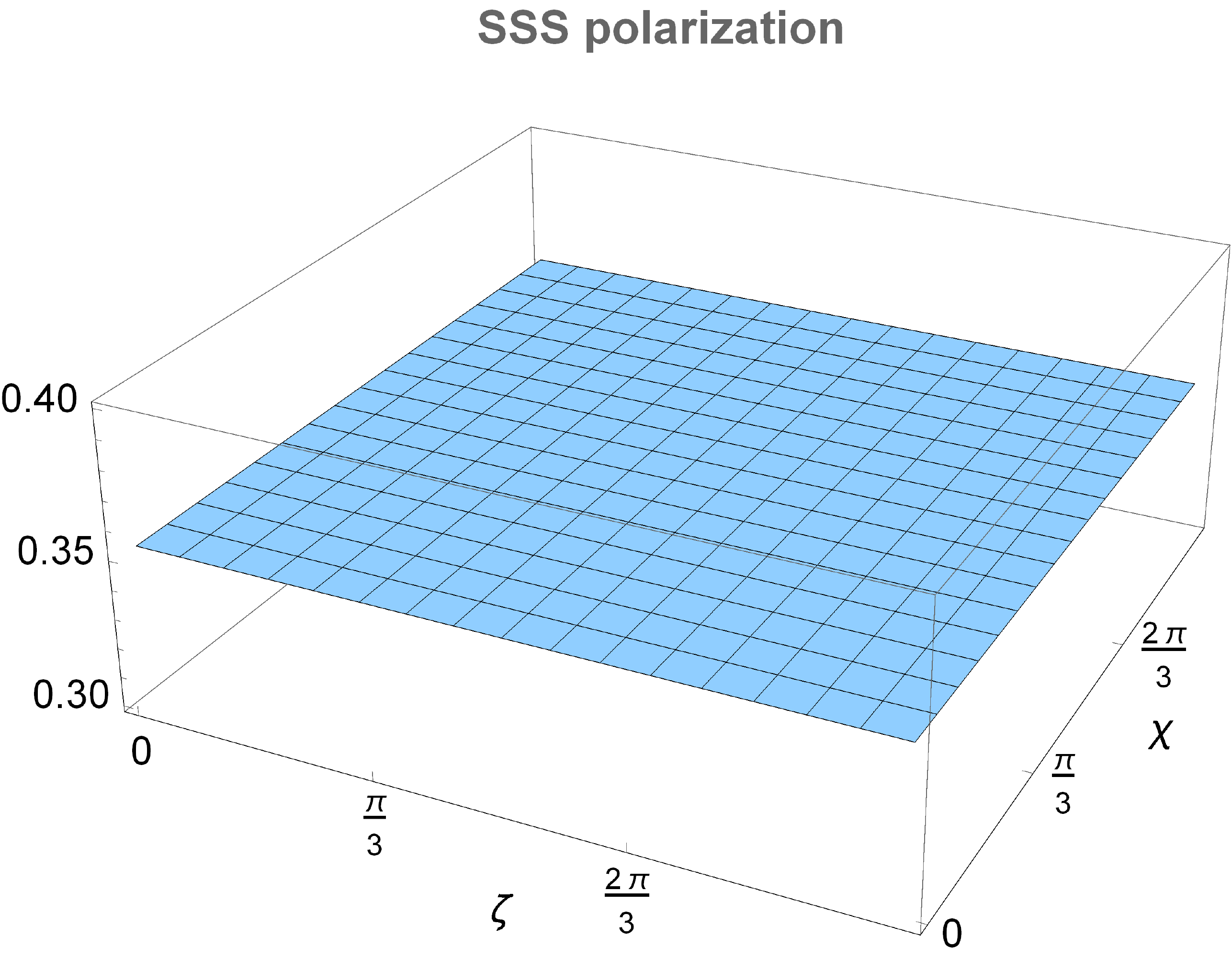}
 \caption{\it Three-point overlap  functions  correlating   scalar and tensor modes for signals from three
 different pulsars lying on orthogonal planes. We vary the angle $\zeta$, $\chi$  between 
 the
 unit vectors from the earth towards the pulsars. We correlate tensor (T) polarization  with scalar (S) polarization. 
}
\label{3dpl0sB} 
\end{center} 
\end{figure}

We then consider correlation of three  signals from three distinct pulsars, with the condition that the
three pulsars lie on orthogonal planes, as done in Section \ref{sub-ortho} in a  purely tensor context. 
 Figure \ref{3dpl0sB} shows our results
for the overlap functions for this case. Notice that for the pulsar configuration we consider
the overlap function for scalar autocorrelations is flat.

\subsection{Three-point overlap functions: correlating PTA and ground-based experiments} \label{corrGB}

\begin{figure}[H]
\begin{center}
\includegraphics[width = 0.3 \textwidth]{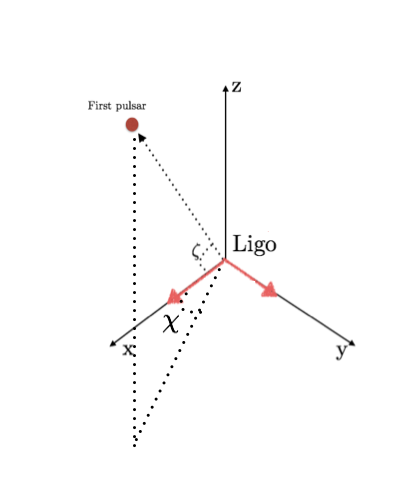}
 \caption{\it Representation of a possible measurement where we correlate signals measured with PTA with signals detected with ground-based experiments (labelled 
 as `LIGO').}
\label{fig:art2} 
\end{center} 
\end{figure}
\begin{figure}[H]
\begin{center}
\includegraphics[width = 0.32 \textwidth]{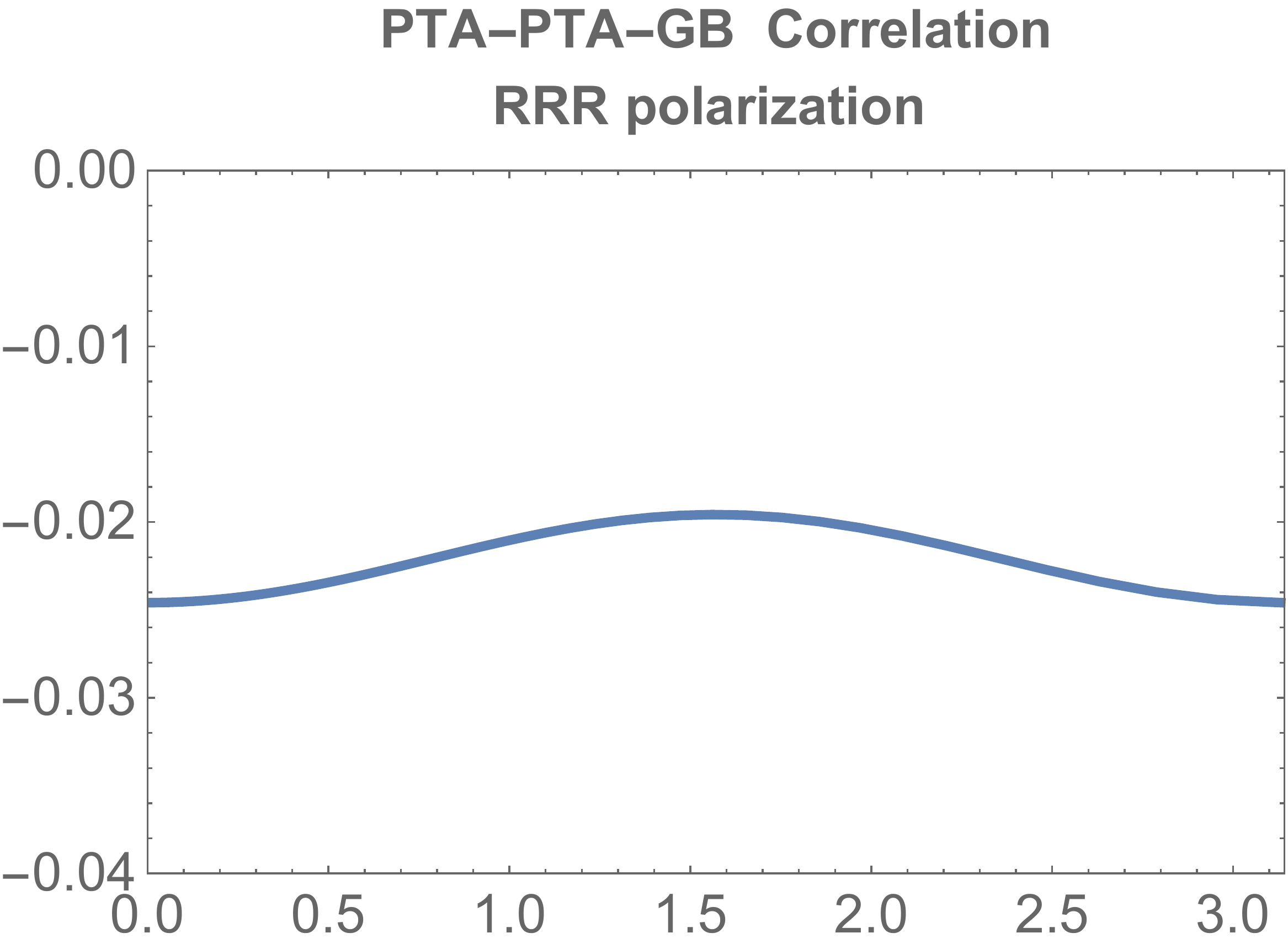}
\includegraphics[width = 0.32 \textwidth]{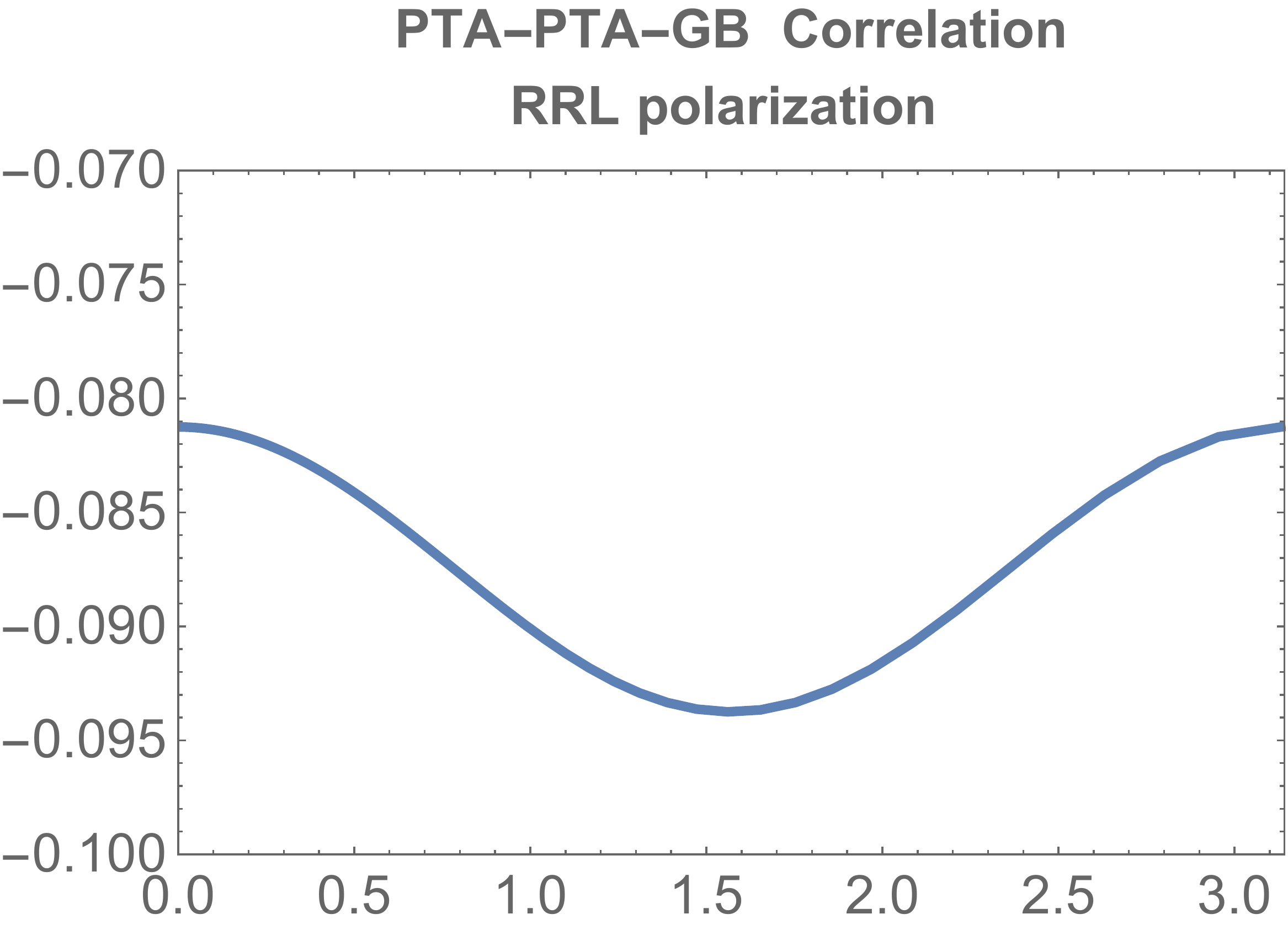}
\includegraphics[width = 0.32 \textwidth]{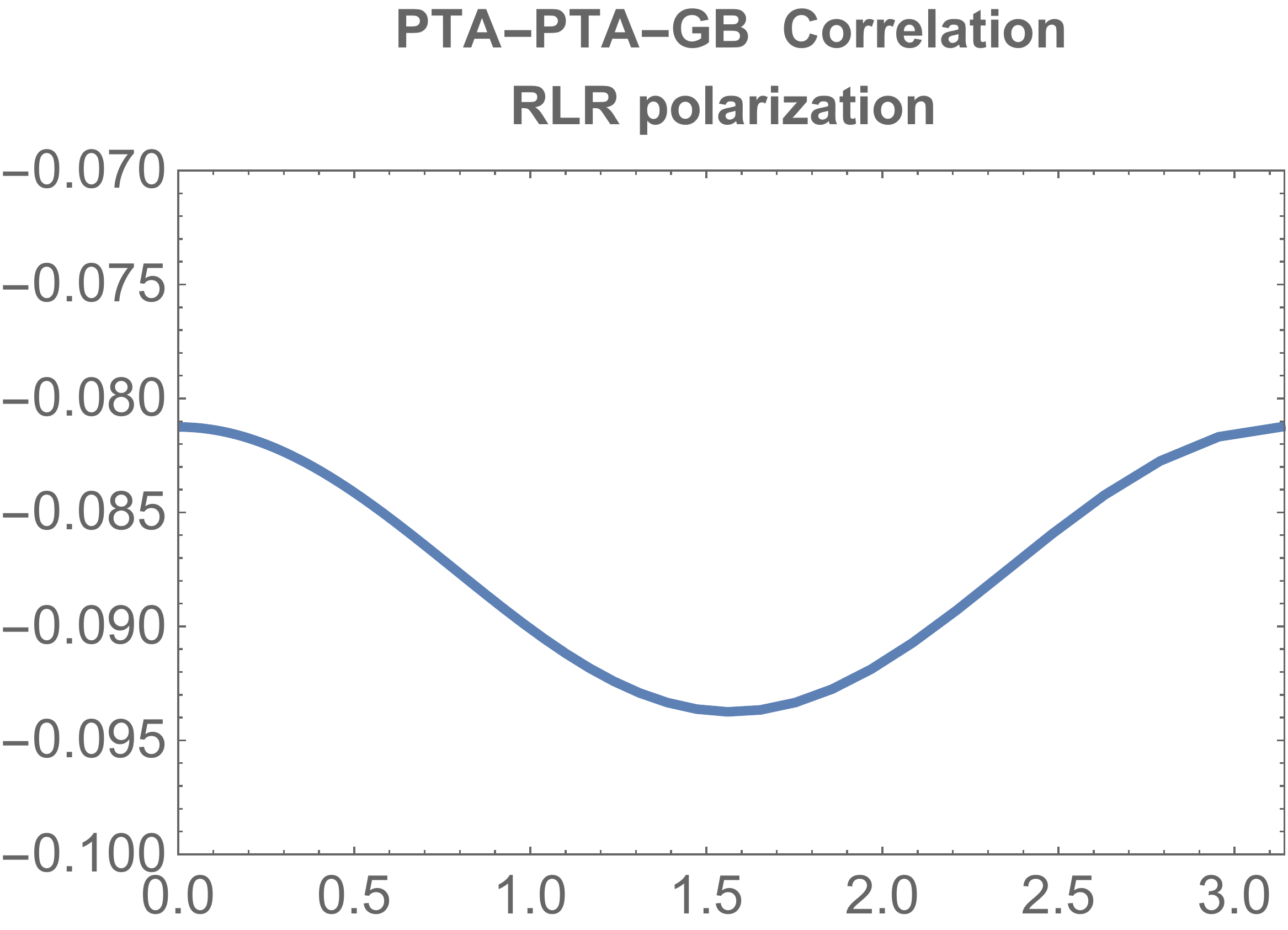}
 \caption{\it  3-point overlap  functions for correlations between two signals from the same pulsar and a signal measured at a  ground-based detector.
 We vary the angle $\zeta$  between 
 the
 unit vectors from the earth and a direction of one of the ground based detector arms.
 %
}
\label{3dpl0} 
\end{center} 
\end{figure}
\begin{figure}[H]
\begin{center}
\includegraphics[width = 0.32 \textwidth]{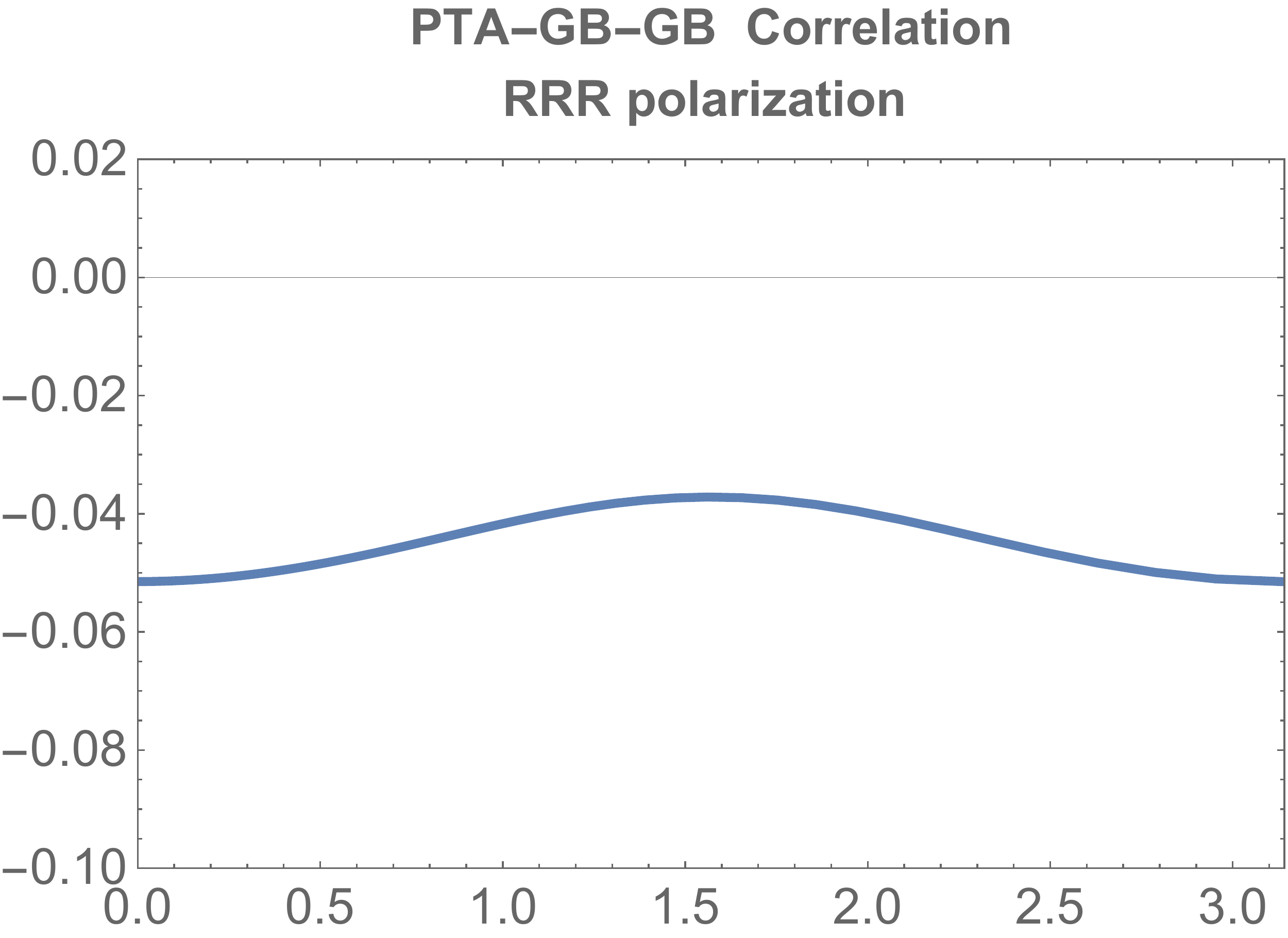}
\includegraphics[width = 0.32 \textwidth]{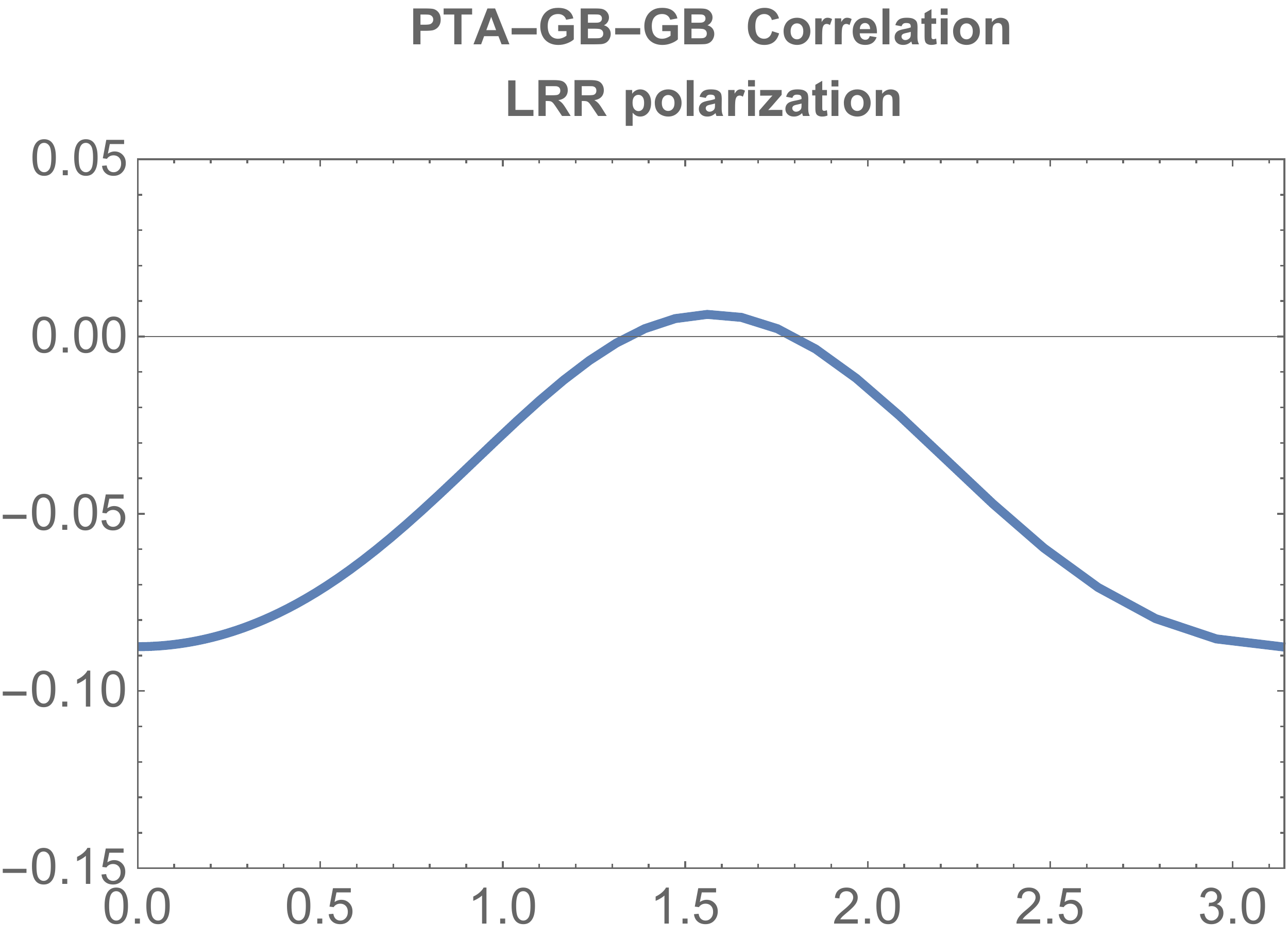}
\includegraphics[width = 0.32 \textwidth]{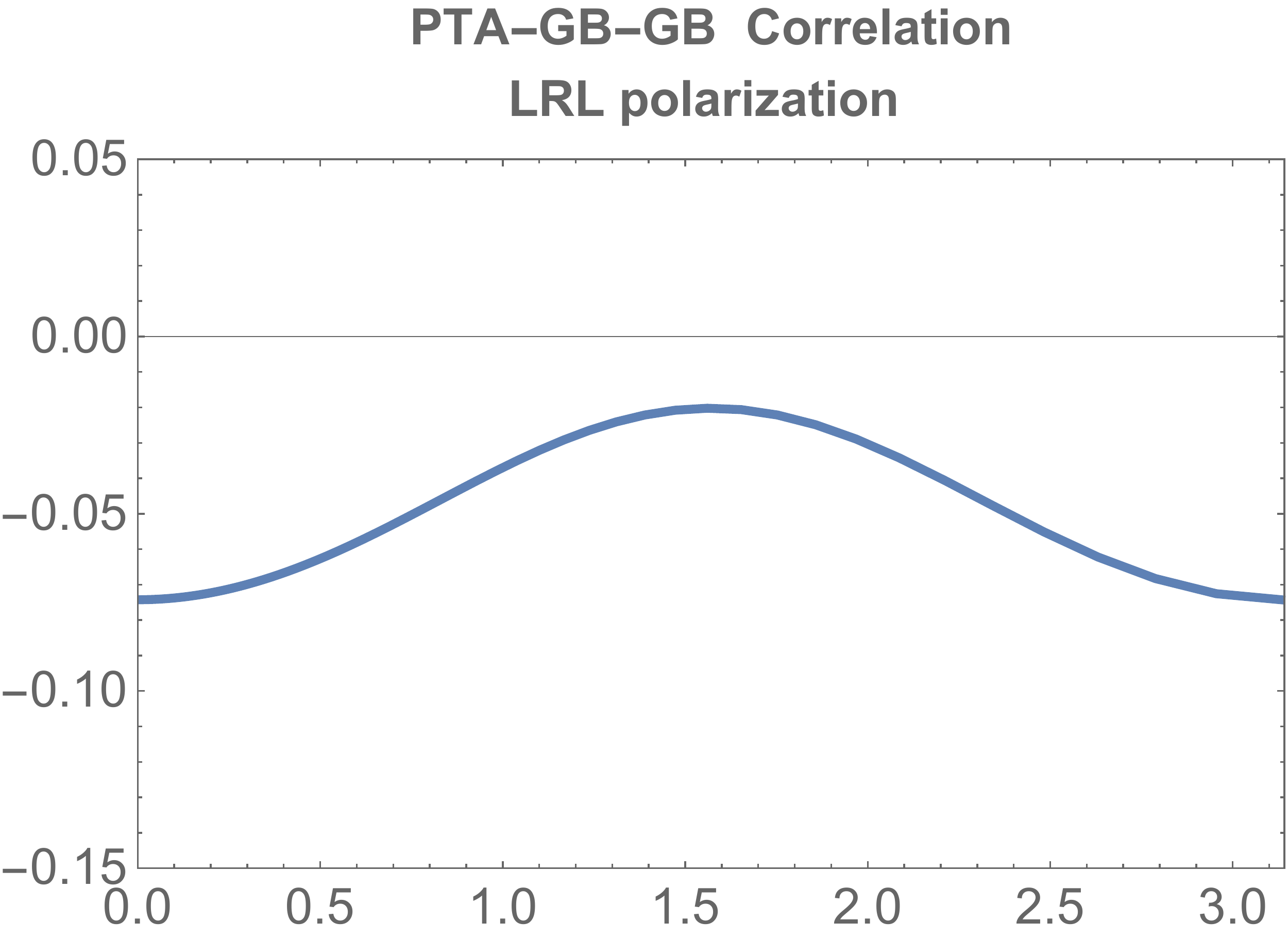}
 \caption{\it   3-point overlap  functions for correlations between one signal from a pulsar and two signals measured at the same  ground-based detector.
 We vary the angle $\zeta$  between 
 the
 unit vectors from the earth and a direction of one of the ground based detector arms.
 %
}
\label{3dpl0a} 
\end{center} 
\end{figure}

Another interesting feature of stationary tensor nG is the possibility 
to correlate signals with very different frequencies, associated with
triangles in momentum space for which one of the side lengths is 
much smaller than the other two (see Fig \ref{fig:folded}, left panel). In cases
in which the  frequencies involved differ by several orders of magnitude, it is interesting
to correlate signals detected by PTA experiments  with
signals measured by ground-based experiments (e.g. LIGO) that work in the frequency range
$10^0-10^3$ Hz.  See Figure \ref{fig:art2} for a graphical representation of the system.
The calculation of the corresponding overlap function is conceptually similar to what we have
done so far: only the expression for the detector tensor (given in eq \eqref{dtPTA} for the PTA case) changes.
In the small-antenna limit, the detector tensor for ground based experiments reads:
\be
F_\alpha^{(\lambda)}(\hat n)\,=\,\frac12\,{\bf e}_{ij}^{(\lambda)}\left(X_a^i\,X_a^j-Y_a^i\,Y_a^j \right)\,,
\ee
with $\vec X_a$, $\vec Y_a$ indicating the arm directions of the interferometer $\alpha$. It is sufficient
to plug this expression for the detector tensor in the general formula \eqref{3ptOv} for the 3-point detector
function  to study the case of cross correlations among different experiments.  
We represent
our results in Figs \ref{3dpl0} and \ref{3dpl0a}. In Fig  \ref{3dpl0}, we correlate two signals
from the same pulsar with a signal measured by a ground-based detector.

 In  Fig  \ref{3dpl0a}, instead,
we correlate one signal from a pulsar with two signals measured by the same ground based detector. 

\section{The optimal signal-to-noise ratio for stationary tensor nG}\label{sec-optimal}

Armed with the results of
Section \ref{sec_PTAoverlap}  on the PTA overlap functions, we now wish to determine the optimal  way to correlate   measurements
from  different pulsars, so to maximise the signal-to-noise ratio (SNR) when measuring stationary
  tensor bispectra in flattened configurations with Pulsar Timing Arrays.  We call $s_\alpha(t)$ the
output of the 
measurement  at time $t$ from pulsar $\alpha$, which is the sum of GW signal $\sigma_\alpha(t)$ and noise $n(t)$. (And same for pulsars $\beta$, $\gamma$.) In order to carry on our computations, we make the following assumptions:
\begin{itemize}
\item We shall assume that the
noise dominates over the
  GW signal. The total time $T$ during which data are collected ($5$-$10$ years) is much larger than the
typical frequency scale of the detected GWs  (say $f\sim 10$ nHz), so
we work in a regime  $f T \gg 1$.  
\item We will assume that the graviton bispectrum is very peaked in  flattened configurations  as described in Section \ref{statio-nonG}, so that
the condition of stationary bispectrum is very well realised,  and any variance associated with the spread of the bispectrum shape
is well smaller than the (dominant) instrumental noise as discussed in the previous bullet point. It would be interesting to study concrete
early universe models leading to graviton nG with this property, but we leave this for future work. 
\end{itemize}

Under these hypothesis,
we start deriving
the expression for the optimal signal-to-noise ratio, which (as we shall see) depends on
the frequency dependence of the tensor bispectrum, as we well as a constant geometrical quantity
that we call 
 ${\texttt r}^{\lambda_1 \lambda_2 \lambda_3}$ ($\lambda_i$ denoting the GW chirality). The quantity  ${\texttt r}^{\lambda_1 \lambda_2 \lambda_3}$ depends on the GW chirality and on
 the number and  geometrical configuration of the pulsars in the 
  PTA system under consideration. Towards the end of the section, we shall collect
 in a table the  values for  this quantity, when evaluated for the IPTA data set \cite{Verbiest:2016vem}.
 
\smallskip 

In order to estimate the optimal SNR, we generalise to the case of 3-point functions
 the  methods described in the textbook \cite{Maggiore:1900zz} and the articles 
  \cite{Anholm:2008wy,Thrane:2013oya}.   The computation
of the optimal SNR for tensor bispectrum has been already carried on in \cite{Bartolo:2018qqn}: here we  simplify the analysis  by adapting the computation to the present
context, where we focus on  `stationary' non-Gaussianity only, and on PTA measurements. 
We define the stationary three point correlator among  a measurement performed with a triplet of pulsars, 
denoted with $(\alpha,\,\beta,\,\gamma)$:
\bea\label{defY}
{\cal Y}_{\alpha \beta \gamma}\,=\,
  \int_{-T/2}^{T/2}\,d t_1 \int_{-T/2}^{T/2}\,d t_2 \int_{-T/2}^{T/2}\,d t_3\,
s_\alpha(t_1)
s_\beta(t_2)
s_\gamma(t_3)
\,Q( t_3-t_1 ) \,Q(t_3-t_2)\,,
\eea
where
  we 
integrate  over the duration $T$ of the experiment.
 In the previous expression,
  $Q(t_i-t_j)$ is a filter function, depending on time differences, which we assume decays very rapidly with increasing values of 
$|t_i-t_j|$. Notice that we include {\it two} copies of $Q$, in order to  take into account the two independent time differences $(t_3-t_1)$ and $(t_2-t_1)$.

In the definition of SNR$\equiv$S$/$N, the quantity S (the signal) corresponds to  the ensemble average value of ${\cal Y}_{\alpha \beta \gamma}$ when the signal is present;  the quantity N (the noise) is the
root mean square value of ${\cal Y}_{\alpha \beta \gamma}$  when the signal is absent. 
In what comes next, we wish to determine the choice of filter function $Q$ maximising the SNR. 
 
We Fourier transform the expression \eqref{defY}:
\bea
{\cal Y}_{\alpha \beta \gamma}&=& 
\int_{-\infty}^{\infty}\,d f_A\,d f_B\,d f_1 \,d f_2\,d f_3\,
\,\delta_T(f_1-f_A)
\,\delta_T(f_2-f_B)
\,\delta_T(f_3+f_A+f_B)
\nonumber
\\
&&
\times
\,\tilde s_\alpha(f_1)
\,\tilde s_\beta(f_2)
\,\tilde s_\gamma(f_3)\,\tilde{Q}(f_A)\,\tilde{Q}(f_B)\,,
\eea
and introduce  the function
 $\delta_T(f)\,\equiv\,\left(\int_{-T/2}^{T/2}\,d t\,\exp{2\pi\,i\,f t}\right)$. This function approaches the Dirac delta function  for $T\to \infty$, and
has the property that  $\delta_T(0)\,=\,T$. 
 We proceed integrating  over the frequencies $f_1$, $f_2$.  We get (in the physically relevant limit of large $f\,T$ where we can approximate $\delta_T$ with a $\delta$ function):
\bea
{\cal Y}_{\alpha \beta \gamma}&=&
\int_{-\infty}^{\infty}\,d f_A\,d f_B\,d f_3\,
\,\delta_T(f_3+f_A+f_B)
\,\tilde s_\alpha(f_A)
\,\tilde s_\beta(f_B)
\,\tilde s_\gamma(f_3)\,\tilde{Q}(f_A)\,\tilde{Q}(f_B)\,.
\eea
We now compute the SNR$\equiv$S$/$N. 

\smallskip
\noindent
{\bf The signal S}. 
Using the procedure outlined above, and equation \eqref{sigFou} for the individual signals
from each pulsar, 
we get the following expression for the total signal associated with the 3-pulsar measurement:
\bea
S&=&\sum_{\lambda_1 \lambda_2 \lambda_3} 
\int_{-\infty}^{\infty}\,d f_A\,d f_B\,d f_3\,d^2 \hat n_1\,d^2 \hat n_2\,d^2 \hat n_3
\,\delta_T(f_3+f_A+f_B)\,F_{\alpha}^{\lambda_1}(\hat n_1)\,F_{\beta}^{\lambda_2}(\hat n_2)\,F_{\gamma}^{\lambda_3}(\hat n_3)
\nonumber
\\
&&
\times
\,\langle h_{\lambda_1}(f_A,\hat n_1)
\, h_{\lambda_2}(f_B, \hat n_2)
\, h_{\lambda_3}(f_3, \hat n_3)
\rangle
\,\tilde{Q}(f_A)\,\tilde{Q}(f_B)\,,
\\
&=&\sum_{\lambda_1 \lambda_2 \lambda_3} \,\int_{-\infty}^{\infty}\,d f_A\,d f_B\,d f_3\,\delta(f_3+f_A+f_B)\,
\delta_T(f_3+f_A+f_B)\,,
\nonumber
\\
&& 
\,\tilde{Q}(f_A)\,\tilde{Q}(f_B)\,{ B}^{\lambda_1 \lambda_2 \lambda_3}(f_A, f_B,\,\hat n_\star)\,{\cal R}_{\alpha \beta\gamma}^{\lambda_1 \lambda_2 \lambda_3}(\hat n_\star)\,,
\\
&=&T\,\sum_{\lambda_1 \lambda_2 \lambda_3} \,\int_{-\infty}^{\infty}\,d f_A\,d f_B\,,
\,\tilde{Q}(f_A)\,\tilde{Q}(f_B)\,{ B}^{\lambda_1 \lambda_2 \lambda_3}(f_A, f_B,\,\hat n_\star)\,{\cal R}_{\alpha \beta\gamma}^{\lambda_1 \lambda_2 \lambda_3}(\hat n_\star)\,,
\label{resS}
\eea
where we use the fact that $\delta_T(0)=T$, and make use of expression \eqref{builcor3} for the GW three point function in momentum
space.

\smallskip
\noindent
{\bf The noise N}. We  
assume that the noise is Gaussian and  uncorrelated between pulsars; we define the noise spectrum as
\be
  \langle n_{\alpha}(f_A) 
 n_{\alpha}^*(f_C) \rangle\,=\,
S_n\,\delta(f_A-f_C)\,,
\ee
where we use the fact that PTA noise is frequency independent. We make
use of the following formula   from \cite{Thrane:2013oya}:
\be
S_n\,=\,2 \Delta t\,\sigma^2\,,
\ee
with $1/\Delta t$ the measurement cadence (of order $20$ yr$^{-1}$), and $\sigma^2$ the {\it rms} of  the noise timing (for IPTA pulsars, this quantity is of  order  $1$ $\mu$s). 
 The noise squared of the measurement, in absence of signal, is
\bea
N^2&=&\langle {\cal Y}  {\cal Y}^*  \rangle\,,
\\
&=&
 \int d f_A\,d f_B\,d f_3\,d f_C\,d f_D\,d f_4\,\,\delta_T(f_3+f_A+f_B)\,\delta_T(f_4+f_C+f_D)
\nonumber\\
&&\times  \langle n_{\alpha}(f_A) 
 n_{\alpha}^*(f_C) \rangle
 \langle n_{\beta}(f_B) 
 n_{\beta}^*(f_D) \rangle
  \langle n_{\gamma}(f_3) 
 n_{\gamma}^*(f_4) \rangle
 \nonumber\\
&&\times Q(f_A)  Q^*(f_C)  Q(f_B)  Q^*(f_D)\,,
\\
&=&T\,
S_n^3\, \int d f_A\,d f_B\,
|  Q(f_A)  |^2 \,|  Q(f_B) |^2\, \label{resN}\,.
\eea
Then the  ratio corresponding to  signal-to-noise is the ratio of the quantity $S$ in eq \eqref{resS} versus
$N$ of eq \eqref{resN}. We can sum over all the available distinct pulsar triplets $(\alpha,\,\beta,\,\gamma)$
in the network under consideration \footnote{Instead of correlating signals
from three different pulsars, we can also correlate two signals from the same pulsar with a signal from a different pulsar, as described in Section \ref{sec_PTAoverlap}. In this case, we need to sum over distinct pulsar pairs, instead
of triplets (more on this later).}.
We find
\be \label{expSNR}
\text{SNR}\,=\,\sqrt{T}\,\frac{\left[\sum_{\alpha\beta\gamma} \sum_{\lambda_i}
\int_{-\infty}^{\infty}\,d f_A\,d f_B\,
\,\tilde{Q}(f_A)\,\tilde{Q}(f_B)\,\left( \,{ B}^{\lambda_1 \lambda_2 \lambda_3}(f_A, f_B,\,\hat n_\star)\,{\cal R}_{\alpha \beta\gamma}^{\lambda_1 \lambda_2 \lambda_3}(\hat n_\star) \right)
 \right]}{\left[ \,S_n^3\, \int d f_A\,d f_B\,
 |  Q(f_A)  |^2 \,|  Q(f_B) |^2\,
\right]^{1/2}}\,.
\ee
In order to determine the function $Q$ maximising the previous expression, we first
define a positive definite scalar product $[\cdots,\cdots]$ between two quantities which depend on frequency (recall that $S_n$ is a positive quantity):
\be
\left[ A_1( f_a, f_b),\,A_1( f_a, f_b) \right]
\,=\, \int d f_a\,d f_b\,
\,A_1(f_a,  f_b)  \,A_2^*(f_a, f_b)\,S_n^3\,.
\ee
By dropping indexes, the SNR of eq \eqref{expSNR}
can be then  schematically
re-expressed in terms of this scalar product as 
\be
\text{SNR}\,=\,\sqrt{T}\,\frac{\left[Q Q,\,{ B} \,{\cal R}/S_n^3\right]}{\left[Q Q,\,Q Q\right]^{1/2}}\,.
\ee
The previous quantity is maximized by choosing the  function $Q$ as
\be
Q(f_a) Q(f_b)\, =\,\frac{{ B}(f_a, f_b, \,n^*)\,{\cal R}(n^*)}{S_n^3}\,.
\ee
 To summarize,  the optimal SNR is given by the expression
\be
\text{SNR}_{\text{opt}}\,=\,\sqrt{4\,T}\, \left[ \sum_{\lambda_1 \lambda_2 \lambda_3} \,
  \frac{  
\int_{0}^{\infty}\,d f_A\,d f_B
  \left({\texttt r}^{\lambda_1 \lambda_2 \lambda_3}\,{ B}^{\lambda_1 \lambda_2 \lambda_3}(f_A, f_B,\,\hat n_\star)\,
  \right)^2}{S_n^3}\right]^{\frac12}\,,
\ee
where we find 
 convenient to define a single quantity ${\texttt r}^{\lambda_1 \lambda_2 \lambda_3}$ containing
the sum  of the PTA response functions over all independent pulsar triplets:
\be\label{defrl}
{\texttt r}^{\lambda_1 \lambda_2 \lambda_3}
\,=\,\left[ \sum_{\alpha\beta\gamma}\left(
{\cal R}_{\alpha \beta\gamma}^{\lambda_1 \lambda_2 \lambda_3}(\hat n_\star)  \right)^2\right]^\frac12\,.
\ee

\begin{table}
\begin{center}
\begin{tabular}{ l c c }
\hline
\hline

 \cellcolor[gray]{0.9}  ~~&~~    \cellcolor[gray]{0.9} 
  Case 1  ~~&~~   \cellcolor[gray]{0.9}  Case 2 \\

\hline
${\texttt r}^{R R R  }$ ~~&~~ $20.17$ ~~&~~ $0.58$ \\
${\texttt r}^{R R L  }$ ~~&~~ $19.58$ ~~&~~ $0.99$ \\
${\texttt r}^{R LR  } $ ~~&~~ $19.58$ ~~&~~ $0.99$ \\
%
\hline
${\texttt r}^{STT } $ ~~&~~ $38.11$ ~~&~~ $0.28$ \\
${\texttt r}^{SST } $ ~~&~~ $119.49$ ~~&~~ $1.96$ \\
${\texttt r}^{SSS  } $ ~~&~~ $168.99$ ~~&~~ $9.12$ \\
\hline
\hline
\end{tabular}
\caption{\label{tabpta1} {\it Value of the quantity ${\texttt r}^{\lambda_1 \lambda_2 \lambda_3}$ for different GW tensor polarizations, computed for two
different cases using IPTA data. Case 1: we do the sum of eq \eqref{defrl} summing over distinct pulsar triplets. Case 2: the sum is made
assuming that we correlate two signals from one pulsar with one signal from another pulsar (as in Section \ref{secsame}) hence we sum over
distinct couples of pulsars. We consider purely tensor correlators (depending on chirality, upper
part of the table) and scalar-tensor correlators (lower part of the table). }}
\end{center}			
\end{table}

\bigskip

Hence we learn that the optimal SNR, besides than on the frequency dependence of the bispectrum,  
is characterized by  the constant quantity ${\texttt r}^{\lambda_1 \lambda_2 \lambda_3}$ of   eq. \eqref{defrl}. This  depends  on the GW polarization, and on the number and position of the pulsars one considers. We compute this quantity 
for the case of IPTA pulsars -- an international collaboration monitoring the period of 49 pulsars from different PTA data set, see Appendix
\ref{appIPTA} -- in Table \ref{tabpta1}.
Summing over distinct triplets of pulsars (Case 1) give much larger values for the parameter
$r^{\lambda_1 \lambda_2 \lambda_3}$, than summing
over couples (Case 2): we believe that this is due to the large number of  independent
triples one can form with the large pulsar
 data set we use. Also, both for Case 1 and 2, the  size of this parameter is much larger in the case of  correlations involving scalar modes only,
since the corresponding overlap function is one order of magnitude larger than in the case
of tensor correlators, see Section \ref{sec_PTAoverlap}.

Our conclusion is that the optimal SNR for measuring stationary graviton non-Gaussianity can be greatly enhanced by monitoring larger numbers of
pulsars,  since it depends on number and configurations of pulsar triplets. It will be interesting to  build explicit models leading to stationary
tensor nG, and investigate at what extent we can probe amplitude and slope (i.e. frequency dependence) of a stationary tensor bispectrum
with current (IPTA) and future (SKA) PTA experiments. We leave these investigations to future work.



\section{Conclusions}

In this work we introduced the concept of stationary graviton non-Gaussianity (nG). We discussed how its properties
make it the only type of graviton nG
that can be  directly measurable in terms of three-point functions of a stochastic gravitational wave
background (SGWB). When evaluated in Fourier space, 3-point functions associated with stationary nG correspond
to configurations peaked in folded configurations.
  We determined 3-point  overlap functions 
for probing stationary  nG  with  PTA experiments, and we  obtained  the
 corresponding optimal signal-to-noise ratio (SNR).
  For the first time, we considered 3-point overlap functions for PTA including scalar graviton polarizations (which can be motivated
  in theories of modified gravity); moreover, we also calculated 3-point overlap functions for correlating PTA with ground based
  GW interferometers. 
 We have shown that 
  the value of the optimal SNR depends on the number and position of monitored pulsars. We built  geometrical quantities  characterizing how the SNR  depends on the PTA  system under consideration, and   we  evaluated such geometrical parameters     using  data from      the IPTA collaboration. We  shown  that monitoring a large number of pulsars can increase the SNR associated       with measurements of stationary graviton nG.

\smallskip

If  in the future a SGWB will be detected with PTA GW experiments, it will be interesting to try to measure  the corresponding signal 3-point function with the tools we developed here. If data will provide 
evidence for stationary nG,  a challenge for theorists will be to design and characterize  early universe scenarios (possibly using approaches based on the effective field theory of inflation) able to realize
folded configurations for graviton nG in momentum space. 


\subsection*{Acknowledgments}
Is is a pleasure to thank Enrico Barausse, Nicola Bartolo, Valerio De Luca,  Emanuela Dimastrogiovanni, Matteo Fasiello, Gabriele Franciolini,  Marco Peloso, and Toni Riotto for  useful discussions and suggestions,  and for comments 
on a draft of this work. G.T. is partially supported by STFC grant ST/P00055X/1.

\begin{appendix}
\section{Conventions for the polarization tensors}\label{app:conventions}
We adopt standard conventions for the polarization tensors adapted to the GW direction $\hat n$,
the same used in \cite{Bartolo:2018qqn}. We assume a Cartesian coordinate system with orthogonal axis $(\hat x,\, \hat y, \,\hat z)$.  Starting from the GW direction $\hat n$ we define two orthogonal unit vectors $\hat u$ and $\hat v$:
\bea
\hat u&=&\frac{\hat n\times\hat z}{|\hat n\times \hat z|}\hskip1cm,\hskip1cm
\hat v\,=\,\hat n\times \hat u\,.
\eea
Using these unit vectors, we can define two tensor polarizations $(+,\,\times)$:
\bea
e^{(+)}_{ab}&=&\frac{\hat u_a \hat u_b-\hat v_a \hat v_b}{\sqrt{2}}
\,,
\\
e^{(\times)}_{ab}&=&\frac{\hat u_a \hat v_b+\hat v_a \hat u_b}{\sqrt{2}}
\, 
\eea
which satisy $e^{(+)}_{ab}  e^{(+)}_{ab}\,=\,1\,=\,e^{(\times)}_{ab}e^{(\times)}_{ab}$, $e^{(+)}_{ab}  e^{(\times)}_{ab}\,=\,0$. 
From these objects, we obtain the chiral polarization operators $(R,L)$ which we use in the main text:
\bea
e^{(R,\,L)}_{ab}&=&\frac{ 
e^{(+)}_{ab}\pm i \,e^{(\times)}_{ab}
}{\sqrt{2}}\,.
\eea
For what respect the breathing scalar mode we adopt the following polarization tensor \cite{Nishizawa:2009bf}:
\bea
e^{(S)}_{ab}&=&\frac{\hat u_a \hat u_b+\hat v_a \hat v_b}{\sqrt{2}}\,.
\eea

\section{The IPTA data set}\label{appIPTA}
 Table \ref{tableIPTAs} contains data from 49 pulsars which are observed by the International Pulsar Timing Array (IPTA). The IPTA consists of various pulsar timing arrays throughout the world. This includes the Parkes Pulsar Timing Array (PPTA) in Australia, NanoGrav (consisting of Arecibo (Puerto Rico) and Green Bank Telescope (USA)) and the European Pulsar Timing Array (EPTA) (consisting of Effelsberg Radio Telescope (Germay), Nan\c{c}ay (France), Lovell Telescope (UK), Sardina Radio Telescope (Italy) and Synthesis Radio Telescope (Netherlands)). The combination of these various arrays allows for a larger data set of pulsars to be observed. 

\begin{center}
	\begin{table}[h!]
	\begin{tabular}{||c c c c||} 
		\hline
		$\textbf{Pulsar Name}$ & $\textbf{RMS}\,[\mu s]$ & $\textbf{TOAs}$ &\\ [0.4ex] 
		\hline
		J0030+0451 &  1.9 & 1,030 & \\
		\hline
		J0034-0538 & 4.4 & 267&\\
		\hline
		J0218+4232 & 6.7 & 1,005 &\\
		\hline
		J0437\text{--}4715 & 0.3 & 5,052&\\
		\hline
		J0610\text{--}2100 & 5.2 & 347&\\
		\hline
		J0613\text{--}0200& 1.2 & 2,940&\\
		\hline
		J0621+1002 & 11.5 & 637&\\
		\hline
		J0711\text{--}6830& 2.0& 549&\\
		\hline
		J0751\text{--}6830& 3.5& 1,129&\\
		\hline
		J0900\text{--}3144& 3.4& 575&\\
		\hline
		J1012+5307& 1.7& 2,910&\\
		\hline
		J1022+1001& 2.2& 1375&\\
		\hline
		J1024\text{--}0719& 5.9& 918&\\
		\hline
		J1045\text{--}4509& 3.3& 635&\\
		\hline
		J1455\text{--}3330& 4.0& 1,495&\\
		\hline
		J1600\text{--}3053& 0.8 & 1,697&\\
		\hline
		J1603\text{--}7202& 2.3 & 483&\\
		\hline
		J1640+2224& 2.0& 1,139&\\
		\hline 
		J1643\text{--}1224&2.7&2,395&\\
		\hline
		J1713+0747&0.3&19,972&\\
		\hline
		J1721\text{--}2457&25.5&152&\\
		\hline
		J1730\text{--}2304&2.1&5.63&\\
		\hline
		J1732\text{--}5049&2.5&242&\\
		\hline
		J1738+0333&2.6&206&\\
	    \hline
		J1744\text{--}1134&1.1&2,589&\\
	\hline
		\end{tabular}
\quad
\begin{tabular}{||c c c c||}
\hline
$\textbf{Pulsar Name}$ & $\textbf{RMS}\,[\mu s]$ & $\textbf{TOAs}$ &\\ [0.5ex] 
\hline
		J1751\text{--}2857&2.4&78&\\
     	\hline
		J1801\text{--}1417& 4.6&86&\\
		\hline
		J1802\text{--}2124&4.3&433&\\
		\hline
		J1804\text{--}2717&4.5&76&\\
		\hline
		J1824\text{--}2452A&2.4&298&\\
		\hline
		J1843\text{--}1113&1.7&186&\\
		\hline
		J1853+1303&1.1&566&\\
		\hline
		J1857+0943&1.3&1,641&\\
		\hline
		J1909\text{--}3744&0.2&2,623&\\
		\hline
		J1910+1256&3.0&597&\\
		\hline
		J1911+1347&0.6&45&\\
		\hline
		J1911\text{--}1114&5.3&81&\\
		\hline
		J1918\text{--}0642&1.5&1,522&\\
		\hline
		J1939+2134&70.0&3,905&\\
		\hline
		J1955+2908&5.0&319&\\
		\hline
		J2010\text{--}1323&1.9&296&\\
		\hline
		J2019+2425&8.8&80&\\
		\hline
		J2033+1734&13.3&130&\\
		\hline
		J2124\text{--}3358&3.0&1115&\\
		\hline
		J2129\text{--}5721&1.2&447&\\
		\hline
		J2145\text{--}0750&1.2&2,347&\\
		\hline
		J2229+2643&3.8&234&\\
		\hline
		J2317+1439&1.6&867&\\
		\hline
		J2322+2057&6.9&199&\\
		\hline
	\end{tabular}
\caption{\it Pulsars analyzed by IPTA \cite{Verbiest:2016vem}.}
\end{table}\label{tableIPTAs}
\end{center}

 J1824-2452A is followed by a letter in order to differentiate between pulsars that are in close proximity (otherwise it is difficult to give them unique names). 

 The pulsar names are determined by the coordinates in the sky. Pulsars with a J in front of the coordinates mean that they have more precise coordinates than older pulsars (before 1993 and these are denoted with a B). The first number is the right ascension which is the a point along the celestial equator between the sun and the March equinox to the point above the earth which is in question. This is measured in hours and minutes. The other section of the coordinates is determined by its declination which is how far above (+) or below (\text{--}) the pulsar is with respect to the celestial equator. Note that older pulsars (that begin with a B) only have their declination to the nearest degree whilst the J pulsars have more accurate coordinates. 
\\
\par Right ascension can be converted to degrees as follows:

\begin{equation}
RA_{deg} =\text{hour} +\frac{\text{min}}{60} + \frac{\text{sec}}{3600}.
\end{equation}


\end{appendix}


\addcontentsline{toc}{section}{References}
\bibliographystyle{utphys}

\bibliography{refsPTA}

\end{document}